\documentclass{nature}
\usepackage{graphicx} 
\usepackage{color}
\usepackage{xcolor}
\usepackage{caption}
\usepackage[super,comma,sort&compress]{natbib}

\usepackage[normalem]{ulem} 
\usepackage{amsmath}
\usepackage{amssymb}
\usepackage{float}
\usepackage{rotating}

\title{
Supercooling of the A phase of $^3$He}

\author{Y. Tian$^1$, D. Lotnyk$^1$, A. Eyal$^{1,2}$, K. Zhang$^{3,4}$, N.~Zhelev$^{1}$, T.S.~Abhilash$^{1,5}$, A. Chavez$^{1}$,  E.N.~Smith$^1$,  M. Hindmarsh$^{3,4}$, J.~Saunders$^6$, E. Mueller$^1$, and J.M.~Parpia$^1$
}
\makeatletter 
\let\saved@includegraphics\includegraphics 
\AtBeginDocument{\let\includegraphics\saved@includegraphics} 
\renewenvironment*{figure}{\@float{figure}}{\end@float} 
\makeatother 

\begin{document}
\maketitle

\begin{affiliations}
 \item Department of Physics, Cornell University, Ithaca, NY, 14853, USA
 \item Physics Department, Technion, Haifa, Israel
 \item Department of Physics and Astronomy, University of Sussex, Falmer, Brighton BN1 9QH, UK
 \item Department of Physics and Helsinki Institute of Physics, PL 64, FI-00014 University of Helsinki, Finland
  \item VTT Technical Research Centre of Finland Ltd, Espoo, Finland
 \item Department of Physics, Royal Holloway University of London, Egham, TW20 0EX, Surrey, UK
\end{affiliations}
\date{\today}

\begin{abstract}
\begin{center}
    Abstract:
\end{center}

Because of the extreme purity, lack of disorder, and complex order parameter, the first-order superfluid $^3$He A-B transition is the leading model system for first order transitions in the early universe. Here we report on the path dependence of the supercooling of the A phase over a wide range of pressures below 29.3 bar at nearly zero magnetic field.  The A phase can be cooled significantly below the thermodynamic A-B transition temperature.  While the extent of supercooling is highly  reproducible, it depends strongly upon the cooling trajectory:  The metastability of the A phase is enhanced by transiting through regions where the A phase is more stable.  We  provide evidence that some of the additional supercooling is due to the elimination of B phase seeds formed upon passage through the superfluid transition.
A greater understanding of the physics is essential before the $^3$He can be exploited to model transitions in the early universe.

*Correspondence to: jmp9@cornell.edu
\end{abstract}

\section*{Introduction}

The condensation of $^3$He pairs into a superfluid state occurs via a second order phase transition at a pressure dependent transition temperature, $T_{\rm{c}}$, shown in Figure 1. The anisotropic A phase is  favored at high temperatures and pressures, while the isotropic B phase is the stable phase below the $T_{\rm{AB}}(P)$ line\cite{Greywall86SH,Wiman2016}. In zero magnetic fields the equilibrium phase diagram exhibits a polycritical point\cite{WheatleyPRL1974} (PCP) at which the line of first order transitions ($T_{\rm{AB}}$) intersects the line of second order transitions ($T_{\rm{c}}$) at 21.22 bar and 2.273 mK. The transition between the A and B phases is first order and thus subject to hysteresis. At the PCP, the bulk free energies of the A, B superfluid phases and the normal state are equal.

\begin{figure}
\includegraphics[width=\textwidth]{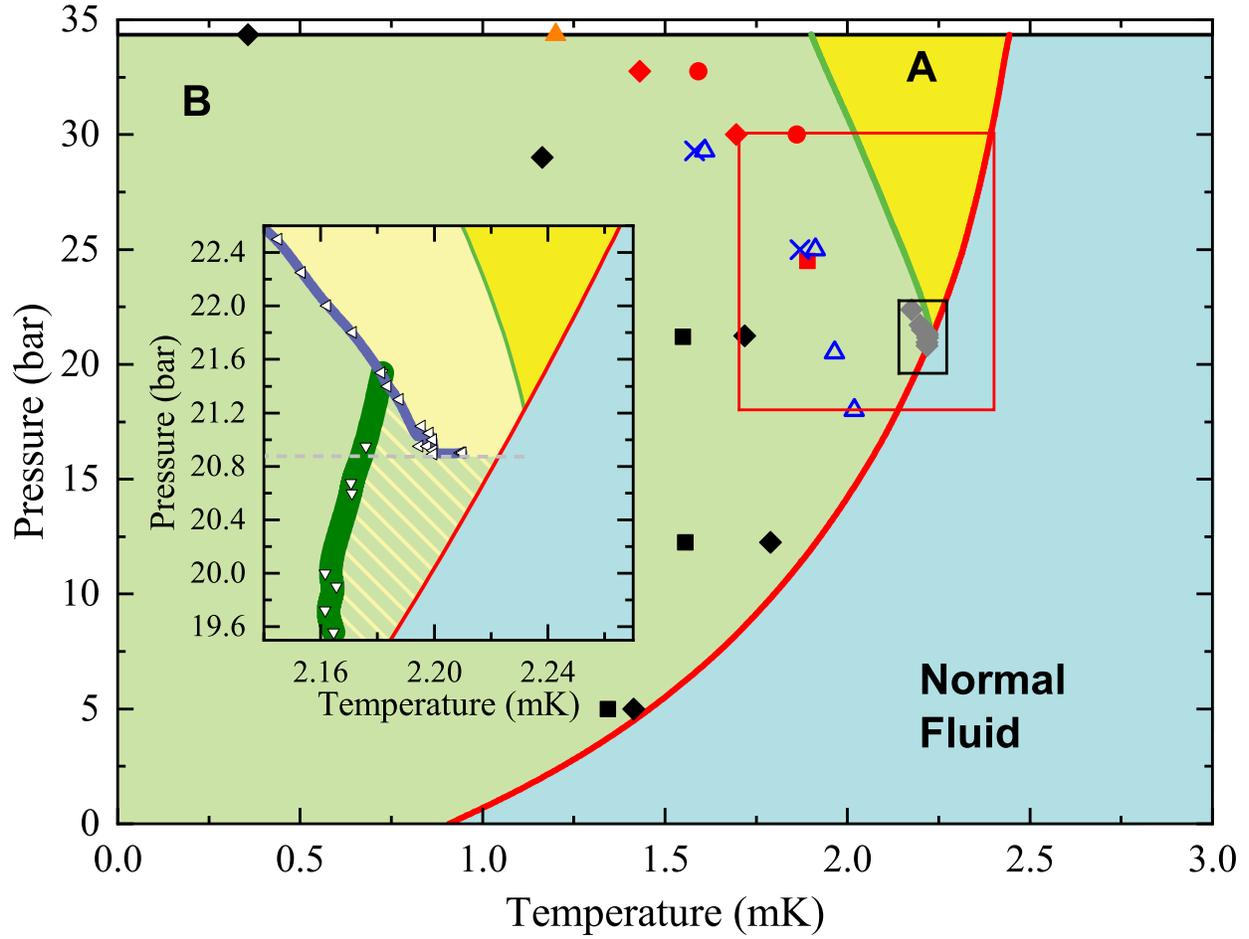} 
\caption{\small
  {\bf Previous Supercooling Results.} The equilibrium phase diagram for superfluid $^3$He. The normal fluid (Blue) is separated from the stable A phase (yellow) and the B phase (green) by a red line that marks $T_c$. The equilibrium A-B transition in zero magnetic field is shown as a green line terminating at the polycritical point (PCP) where the A, B and Normal phases coincide. Centered on the PCP is the region investigated in Ref.~[\citen{LotnykPRL2021}] shown as a black rectangle and also inset. Also shown is the region investigated in this paper (red box). Results from previous investigations elsewhere in a variety of magnetic fields are also shown (\textcolor{gray}{$\blacklozenge$}, 4.9 mT, 0.5 mT \cite{Wheatley1974}; \textcolor{blue}{$\times$}, 56.9 mT; \textcolor{blue}{$\triangle$}, 28.4 mT  \cite{HakonenPRL1985}; \textcolor{orange}{$\blacktriangle$}, 0 mT \cite{Fukuyama1987}; \textcolor{red}{$\boldsymbol{\bullet}$}, 0 mT; \textcolor{red}{$\blacklozenge$}, 10.0 mT; \textcolor{red}{$\blacksquare$}, 20.0 mT \cite{Swift1987};   $\blacksquare$, $\blacklozenge$, 28.2 mT \cite{SchifferPRL1992}.
  }
\label{fig::1_PhaseDiagrama}
\end{figure}

The A phase is highly metastable, and can persist down to extremely low temperatures for 
long times ($\ge$ 1 day) at high pressures, providing surfaces of the container are smooth\cite{OsheroffPRL1992}.
Standard homogeneous nucleation theory\cite{Cahn-Hilliard1958,Langer:1969bc} 
argues that the transition from metastable A to stable B is mediated by thermal fluctuations that produce bubbles of characteristic size $r$.   For small bubbles (size less than the critical radius, $R_{\rm crit}$), the interfacial energy cost ($\propto r^2$) is larger than the bulk free energy gain ($\propto r^3$), but for large bubbles the opposite holds.  Thus, if thermal fluctuations create a bubble with $r<R_{\rm crit}$, it rapidly shrinks.  Conversely, a bubble with $r>R_{\rm crit}$ will grow.
This model\cite{OsheroffPRL1977}, applied to $^3$He, leads to $R_{\rm crit}$ $\approx$ 1.5 $\mu$m, and an activation energy which is many orders of magnitude above the thermal energy\cite{kaul1980surface,Bailin:1980ny,PhysRevLett.53.1096} implying an unobservably small nucleation rate. 
Surface defects potentially alter the energetics (most surfaces favor the A phase\cite{Ambegaokar74} and there is no clean explanation of how they would mediate the A-B transition). Despite extensive experimental\cite{Wheatley1974,HakonenPRL1985,Wheatley1986,BoydSwift1990,OsheroffPRL1992,BauerleNature1996,BunkovPRL1998,BartkowiakPRL2000,LotnykPRL2021} (Figure  \ref{fig::1_PhaseDiagrama}) and theoretical investigations\cite{LeggettResp1985,LeggettPRL1986,LeggettYip1990,LeggettJLTP,HongJLTP,TyePRB2011}, the 
mechanism for B phase nucleation remains a mystery.

Laboratory studies of the dynamics of first order phase transitions  have cosmological implications. Importantly, the possibility of first order
symmetry-breaking phase transitions \cite{Kirzhnits:1972iw,Kirzhnits:1972ut} in the early universe have been used to explain baryon asymmetry\cite{Kuzmin:1985mm}.  The same physics also produces
gravitational waves\cite{Witten:1984rs,1986MNRAS.218..629H,Mazumdar:2018dfl,Hindmarsh:2020hop} that are thought to be detectable in future space based detectors such as Laser Interferometer Space Antenna (LISA) \cite{audley,Caprini:2019egz}. Experimental confirmation of its applicability to a laboratory system would lend more weight to the calculations of gravitational wave production for LISA and other future probes of the early Universe. However, the theory of first order phase transitions in the early Universe\cite{ColemanPRD1977,Linde:1981zj} is based on  the same homogeneous nucleation theory which fails to explain the behavior of $^3$He.

\begin{figure}
\includegraphics[width=\textwidth]{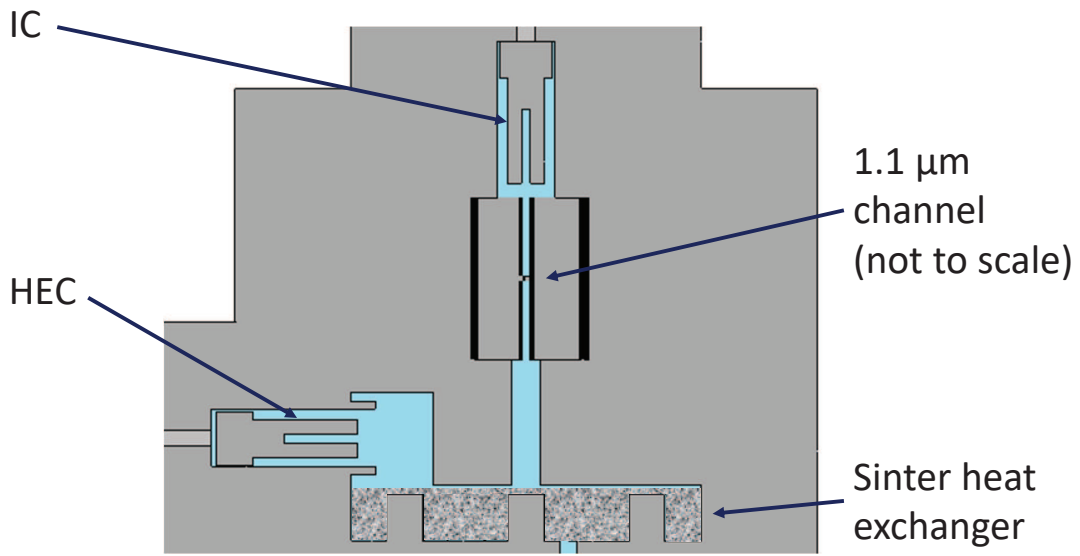} 
\caption{
  {\bf Schematic of Cell.} The Isolated Chamber (IC) and Heat Exchange Chamber (HEC) both contain quartz forks whose quality factor is monitored to determine the phase of the $^3$He superfluid. The chambers are separated by a 100 $\mu$m long channel with aperture 1.1 $\mu$m tall and 3 mm wide.   }
\label{fig::2_CellSchematic}
\end{figure}

Here we study the nucleation of B phase in a pair of chambers connected by a high aspect-ratio ``letterbox" channel
(Figure~\ref{fig::2_CellSchematic}). 
Both the geometry and the surface qualities are relevant:
The smaller chamber (denoted the Isolated Chamber (IC)) incorporates ``ordinary" as-machined coin silver surfaces. It also houses a quartz fork whose resonant properties (frequency and quality factor, $Q$) allow us to infer the phase of the $^3$He in the chamber. The IC is separated from a larger chamber  containing sintered silver by a micromachined channel construction consisting of a 1.1 $\mu$m tall $\times$ 3 mm wide $\times$ 100 $\mu$m long channel and 200 $\mu$m tall $\times$ 3 mm wide  $\times$ 2.5 mm long lead-in channels 
on either side. This construction was nanofabricated in silicon and capped with glass\cite{Zhelev18RSI} (See also Supplementary Note 1). The silver sinter containing chamber (denoted the Heat Exchange Chamber (HEC)) incorporates a quartz fork similar to that contained in the IC.  The A phase is stabilized in confined spaces, and the narrow channel potentially prevents the propagation of an A-B phase boundary from one chamber to the other -- allowing the transitions to be independent.

In an earlier publication\cite{LotnykPRL2021}, we reported initial observations of the reproducibility of B phase nucleation and an  unexpected path dependence for the A phase's stability. From those experiments it was unclear whether the path dependence was limited to the region near the PCP and there were few clues about the microscopic origin of the phenomenon. Here, we have expanded the region investigated to include the highest pressure readily accessible to us (the pressure of the minimum of the $^3$He melting curve, 29.3 bar) and have designed a series of protocols which provide significantly more clarity about the phenomenon.

As already emphasized, homogeneous nucleation theory is unable to explain the nucleation of the B phase from the A phase:  There is a vanishing probability that thermal fluctuations produce a B phase bubble which is larger than the critical radius.  The transition can be triggered\cite{SchifferPRL1992,BauerleNature1996,RuutuNature1996} by bringing a radioactive source near $^3$He -- which is consistent with models where energetic particles (either deliberately introduced or due to Cosmic rays) are responsible for the observed A-B phase transition\cite{PhysRevLett.53.1096,BauerleNature1996,RuutuNature1996}.  Those models can not explain why when $^3$He is repeatedly cooled\cite{HakonenPRL1985,LotnykPRL2021}, the transition consistently occurs along the same temperature and pressure line (dubbed the catastrophe line\cite{HakonenPRL1985}).
Alternate models such as Q balls\cite{HongJLTP} or Resonant Tunneling (RT)\cite{TyePRB2011} have been proposed to explain this behavior.   Those field-theory inspired models, however, are not consistent with the path dependence seen in our experiments.

In an attempt to explain our observations, we note that the silver sinter contains a large number of chambers which are connected to the bulk fluid by narrow channels or constrictions.  
We hypothesise that, upon traversing $T_c$, the B phase is formed in some of these chambers, while the A phase is formed in the bulk. For computational expediency we specify these chambers to be filled with B phase. However, confinement would result in a distorted order parameter quite different from the bulk B phase. Surface tension stabilizes the requisite A-B domain walls at sufficiently small constrictions.  The size of the largest stable domain wall depends on pressure and temperature:  In the A region of the phase diagram, the A phase will rush in to any of the chambers whose opening is larger than this size.  Conversely, the (path dependent) catastrophe line will be determined by the size of the smallest constriction that connects to a B-filled chamber.  This model is similar to the {\em lobster-pot} scenario which was proposed for understanding the nucleation of the A phase from B\cite{LeggettYip1990}.

Cavities in the sinter are unable to explain all of our observations, and it is likely that some other mechanism is also at play.  For example, the A phase  order parameter (in standard experimental geometries) may contain complicated textures with highly frustrated points that may act as seeds for the B phase. Such seeds may involve B-inclusions, or just precursor regions where the A order parameter is strongly suppressed. While some of this structure forms spontaneously due to the Kibble-Zurek mechanism\cite{Kibble1976,Zurek1985}, much of the spatial complexity is likely due to surface effects:  Surfaces constrain the orientation of the order parameter and surface corrugations or scratches can lead to complicated disgyrations and other structures\cite{Maki_JLTP1978}, {\bf perhaps} containing seeds of B phase. Similar to the cavity scenario, the observed A-B transition is set by the size of the ``largest" seed, whose catastrophe temperature is highest. These largest seeds are also the most fragile and may be eliminated by exposure to high pressures where the A phase is most stable. The key feature of the the path-dependence in both scenarios is which seeds survive.
Development of an understanding of this pressure dependence is essential if $^3$He is to to be a useful model for phase transitions in the early Universe.

{ We emphasize that the order parameter of helium is contained in a high dimensional space, and the paths connecting the A and B phases are strongly influenced by surfaces, textures, and distortions from confinement.  Nucleation can occur through both thermal fluctuations and quantum tunneling, the latter of which can display interference effects which are particularly sensitive to such changes in the energy landscape\cite{ColemanPRD1977}.  Models of nucleation in inhomogeneous settings contain a multitude of complexities\cite{chaikin}.}

\section*{Results}
\subsection{Experimental details.}

The normal-superfluid and A-B transitions were detected using quartz forks located in the IC and HEC. The temperatures were obtained with reference to a $^3$He melting curve thermometer\cite{Greywall84TC} mounted on the cold plate of the nuclear demagnetization stage. For details of the operation of the forks and of the thermometry, we refer to the methods section. 

\subsection{Supercooling at constant pressure.}
\begin{figure}
\centering
\includegraphics[width=0.49\textwidth]{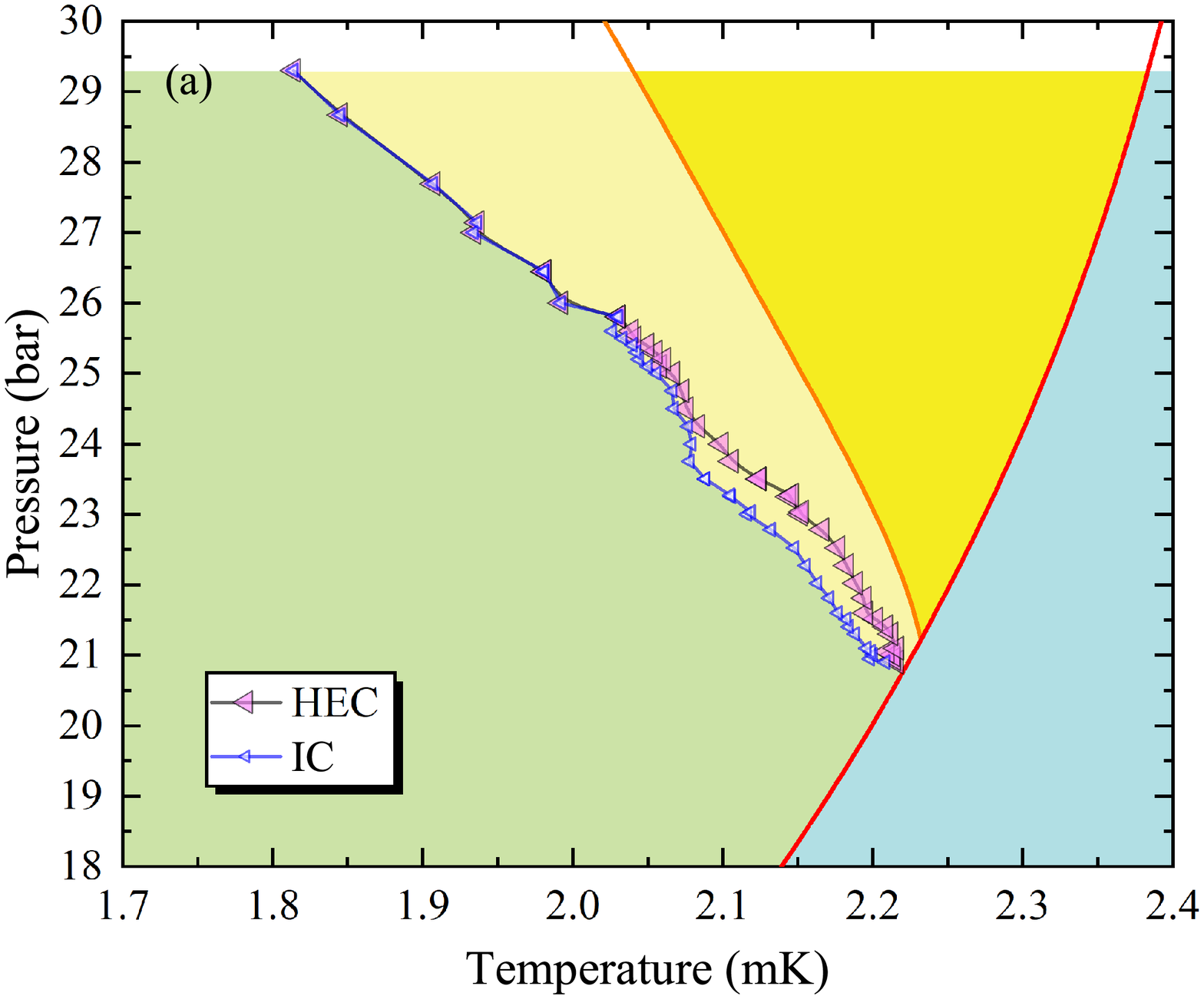}
\includegraphics[width=0.49\textwidth]{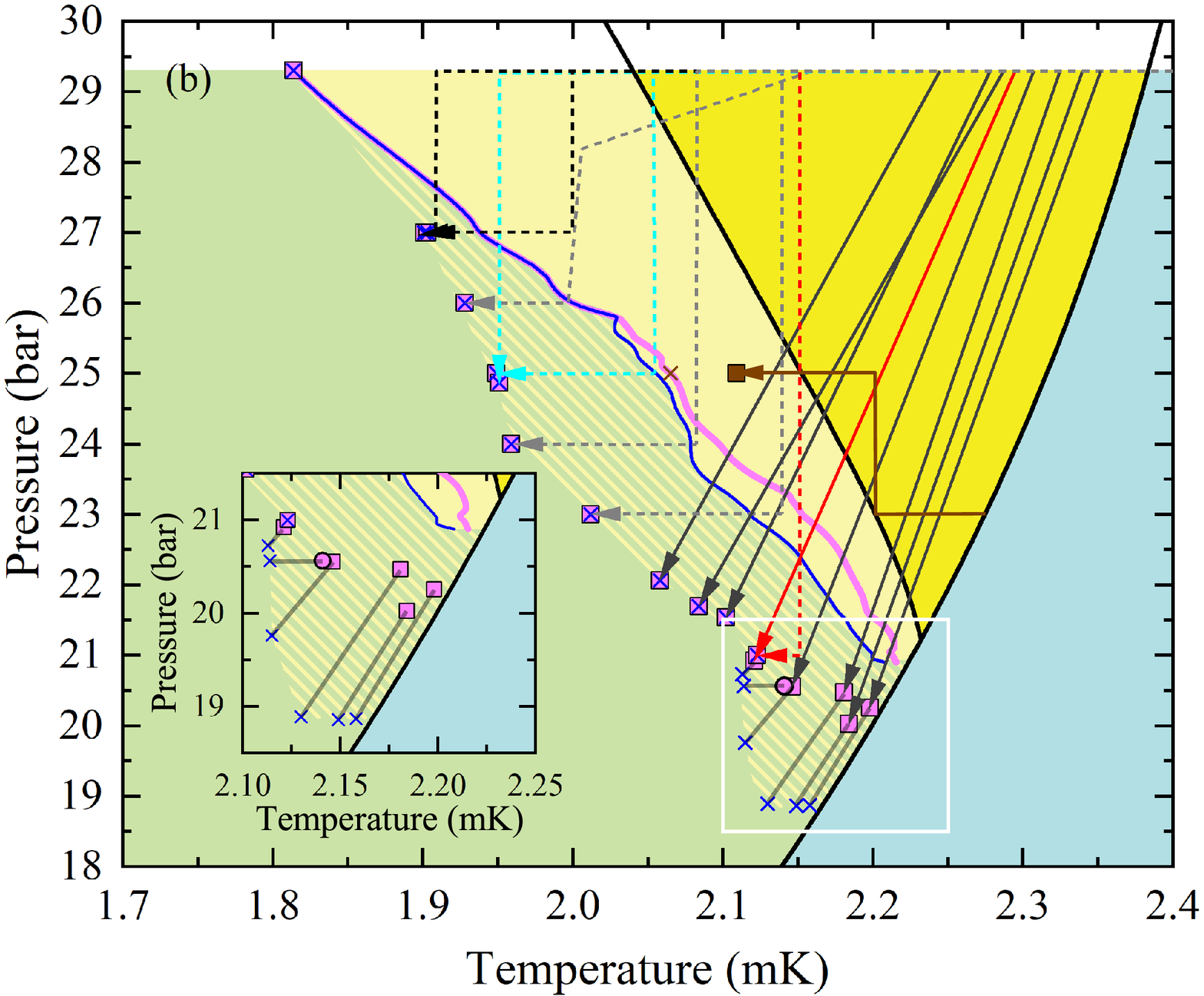}
\caption{\small
  {\bf Constant Pressure and Pressure Varied A-B Transitions.} (a) Constant pressure cooled A-B transitions are shown with pink left pointing symbols denoting transitions in the HEC, blue open left pointing triangles denoting transitions in the IC. Above 25.8 bar the transitions are coincident in time and are shown as nested triangles. (b) A-B transitions observed while decreasing the pressure starting from 29.3 bar are shown along with their paths. Pink Squares show transitions in the HEC, blue crosses transitions in the IC. Where these transitions occur simultaneously, they are superposed. At low pressure they separate with the A-B transition in the IC observed at a lower $P,T$ than that in the HEC (see Inset). Constant pressure cooled transitions from Fig 3 (a) are shown as solid lines. Cyan, black and red lines each show two different paths terminating at the same ($P,T$) coordinates. The brown line shows the result of pressurization followed by further cooling at constant pressure. 
  }
\label{fig::3_ConstantPressure}
\end{figure}

The first set of measurements were carried out while cooling at a constant rate ($\leq$10 $\mu$K/hr) and fixed pressure.   
Figure~\ref{fig::3_ConstantPressure}(a) shows the temperatures at which the A-B phase transition was detected in the HEC (pink triangles) and the IC (blue triangles).
Below 23.8 bar the HEC transition occurs at a substantially higher temperature than in the IC.  In this regime we believe that the A phase is stable in the channel:  It acts as a plug, preventing the A-B wave-front from propogating from the HEC to the IC.  The silver sinter in the HEC leads to more complicated variations of the order parameter, and it is reasonable that the HEC and IC contain different B phase seeds with different catastrophe temperatures.
Between 23.8 bar and 26 bar there is a decrease in the separation between the two transitions, which suggests that the A-B wave-front is only weakly pinned by the channel.  Above 26 bar the two transitions happen simultaneously, and we conclude that in this regime the channel is unable to sustain an A-B interface once the transition is initiated in the HEC. 
(see the Discussion section and Supplementary Notes 2 and 3).

\subsection{Supercooling after decreasing pressure.}

In our model the largest degree of supercooling should occur for trajectories passing through the regions 
where the A phase is most stable ({\it ie.} at high pressure).  
To explore this feature, we first cool at our highest accessible pressure (29.3 bar) followed by depressurizing and further cooling.  In Figure~\ref{fig::3_ConstantPressure}(b) we illustrate several such trajectories.  The solid black lines show cooling trajectories where we maintained an approximately constant $Q$ of the fork in the IC.  This constant-$Q$ condition yields a path which is roughly parallel to $T_c$.  For these trajectories we depressurized by 4-6 bar during the first day, followed by proceeding at 1.3 bar per day, or less.  During the rapid part of the ramp, but not during the slow part, there was some viscous heating observed in the IC.

We found that the extent of supercooling was significantly greater than what we achieved
while cooling at constant pressure (denoted by pink (HEC) and blue (IC) lines instead of data points in Figure~\ref{fig::3_ConstantPressure}(b)). With the exception of the four lowest constant $Q$ runs (closest to $T_c$), the A-B transitions occurred simultaneously in both chambers, and are depicted in  Figure~\ref{fig::3_ConstantPressure}(b) as co-incident crosses and squares. The same symbols (crosses  for IC  and squares for HEC) are used to denote the observed $T,P$ coordinates of the pressure varied  transitions for the four lowest points.  A temperature correction is applied to the IC data to account for thermal offsets between the chambers.

To further explore the path dependence, we considered the trajectories shown as dashed lines in Figure~\ref{fig::3_ConstantPressure}(b). These begin with constant-pressure cooling at 29.3 bar, followed by fixed temperature depressurizations and fixed pressure cooling.  In all cases we observe significantly more supercooling than in Figure~\ref{fig::3_ConstantPressure}(a).  Crucially, there appears to be a definitive locus of points in the $T-P$ plane on which all of the trajectories fall.  As illustrated by the dotted black and cyan paths (terminating at 25 and 27 bar), one finds the same A-B transition points when cooling after depressurizing or depressurizing at constant temperature -- as long as the trajectory passed through the A phase at 29.3 bar. 

In Supplementary Note 4 we present a detailed comparison between the 23 bar fixed pressure run,  and one of the trajectories passing through 29.3 bar before cooling at 23 bar.  We find that the only detectable difference is the temperature of the A-B  transition.  There are no signs of thermal gradients, viscous heating, or other systematic effects. 

\subsection{Other supercooling results.}

\begin{figure}
\centering
\includegraphics[width=\textwidth]{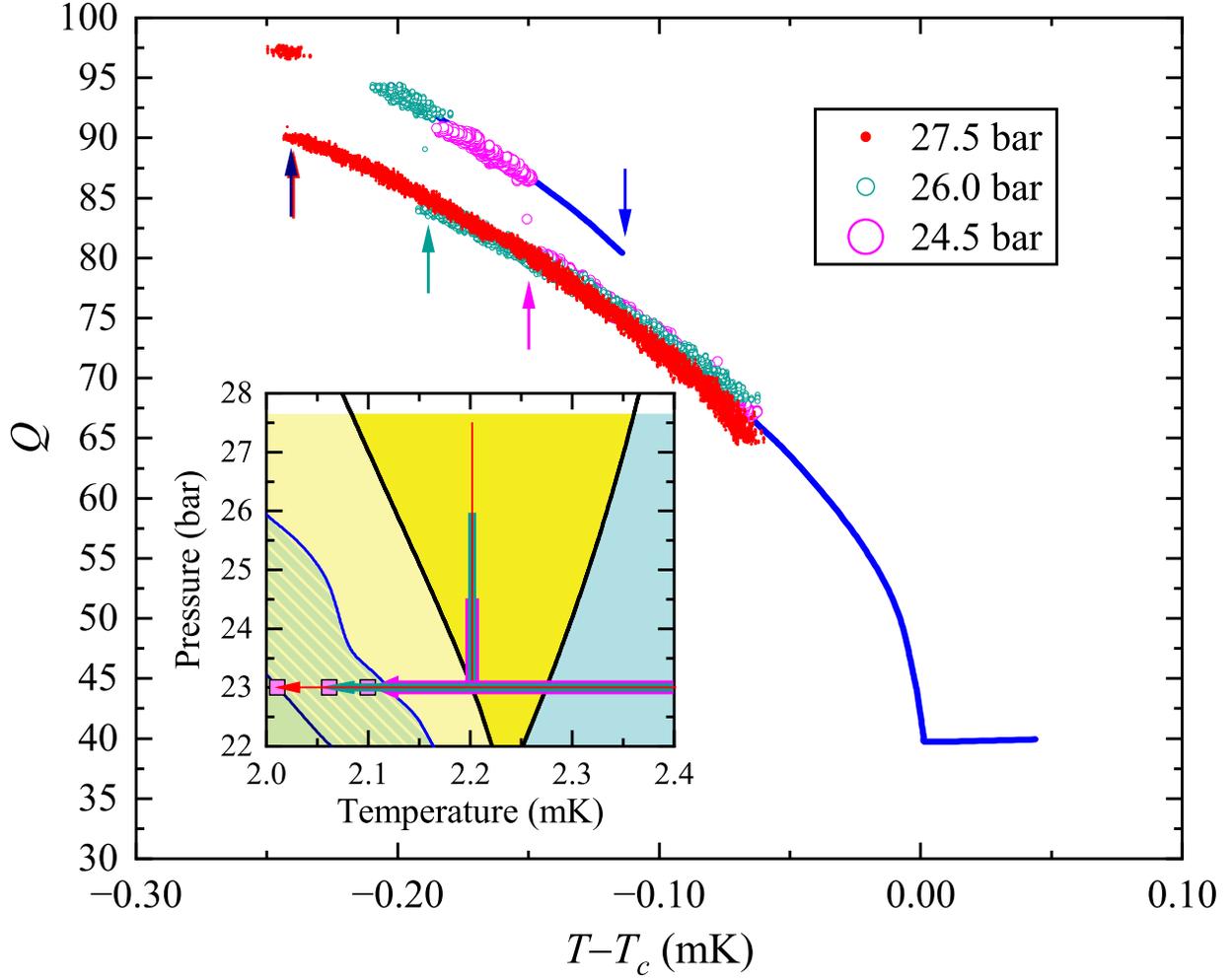} 
\caption{\small
   {\bf Comparison of Constant Pressure and Pressure-Cycled Runs - $Q_{HEC}$ {\it vs} $T$ cooled through $T_c$ and $T_{AB}$ at 23 bar.} The constant pressure cooled experiment in the HEC is shown as a solid blue line. For each of the three pressure cycled runs, after cooling through $T_c$, while the temperature was maintained at $\approx$ 2.2 mK, the $^3$He was pressurized to 24.5 bar (purple), 26 bar (green) and 27.5 bar (pink), then depressurized to 23 bar, and then cooled further at constant pressure till the A-B transition was observed. The HEC and IC transitions were simultaneous for the 26 and 27.5 bar runs. Arrows mark the positions of the various A-B transitions. The inset shows the A-B transitions and the paths in the $P,T$ diagram. The hatched region in the inset is the same as in Figure 3(b).
  }
\label{fig::5_QvsT}
\end{figure}

 Figure~\ref{fig::5_QvsT} illustrates four additional runs, each of which involve cooling at 23 bar.  The blue curve shows the quality factor of the quartz oscillator in the HEC during cooling.  It jumps discontinuously at $T=2.12$ mK ($T-T_c$ = $-0.113$ mK), indicating the A-B phase transition.  For the other three runs, the helium is cooled to 2.2 mK, and then slowly pressurized to $p_{\rm max}=$ 24.5 bar, 25 bar, or 27.5 bar.  The pressure is then reduced back to 23 bar, and the temperature is reduced further.  As expected from our model, the degree of supercooling is a monotonic function of $p_{\rm max}$:  The B phase seeds are suppressed by excursions deep into the equilibrium A phase.  In  these regions the free energy differences between the A and B phases are largest compared to the thermodynamics barriers.  Note, the changes caused by these excursions are subtle enough that they do not appreciably change the quality factors (aside from shifting the A-B transition).

The brown path in Figure~\ref{fig::3_ConstantPressure}(b)
illustrates the reverse effect.  We traverse the stable A phase at 23 bar.  We then increase the pressure to 26 bar before continuing to cool.  We find that the A-B transition occurs at a higher temperature than if we simply cool at 26 bar.  This path avoids the regions of the phase diagram where the A phase is most stable.

To ensure that the supercooling is not significantly affected by the sweep rate, we repeated the experiment in Figure~\ref{fig::5_QvsT}, varying  the rate of  pressurizing and depressurizing during the jog from 23 bar to 27 bar and back. We varied this rate from 1.3 bar/day to 27.5 bar/day, finding no difference in the degree of supercooling after completing the cooling at 10 $\mu$K/hr.

To verify the stability of the A phase obtained after depressurization, we selected a trajectory that terminated below the PCP from Figure~\ref{fig::3_ConstantPressure}(b).  After cooling through $T_c$, starting from 29.3 bar and 2.15 mK we depressurized (at fixed $Q$) to 20.5 bar and stopped at a point within 3 $\mu$K of the the temperature where we previously observed the transition.  We waited at this $T,P$ for 1 day. We then slowly cooled at a rate of 0.5 $\mu$K/hr until we observed the transition in the HEC approximately 2 $\mu$K below the previously observed result (open pink circle in Figure~\ref{fig::3_ConstantPressure}(b)).  Thus the supercooled A phase has a lifetime in excess of 24 hours, and any dynamics which happen on this timescale do not appear to significantly influence the catastrophe line. Further, the A-B transition in the IC ($\times$) occurred at a lower temperature than the transition in the HEC, consistent with the other depressurization runs terminating in this part of the phase diagram (see Figure~\ref{fig::3_ConstantPressure}(b) and its inset).

\section*{Discussion}

We analyze our data by considering the model from our introduction, where the B phase grows from seeds which are contained in small chambers with a distribution of narrow necks. While cooling through the A phase, the A phase intrudes on the chambers with the largest openings:  The size of the remaining B seeds determines the path-dependent location of the observed A-B transition.  Similar logic should apply to the cases where the B seeds are at the nodes of frustrated textures or distortions.

In order to balance forces, an equilibrium domain wall between the A and B phase must be bowed with a mean curvature $\kappa=|\delta f|/(2\sigma)$,  where $\sigma$ is the surface tension and $\delta f$ is the difference in free energy densities between the two phases.  As detailed in the Supplementary Note 3, a circular hole in a flat plate with diameter $W$ will prevent the intrusion of the A phase if $W<\sin(\theta_{A})/\kappa_{A} $, where the contact angle $\theta_{A}$ is determined by surface energies.  Conversely, in the B region of the phase diagram, the same orifice will prevent the B phase from exiting if $W<1/\kappa_{B}$ -- the contact angle does not appear in this expression because the B phase typically does not wet a surface.  Note: these equations are sensitive to our modelling of the geometry of the orifice, and the contact angle depends on the surface properties.


\begin{figure}
\centering
\includegraphics[width=0.8\textwidth]{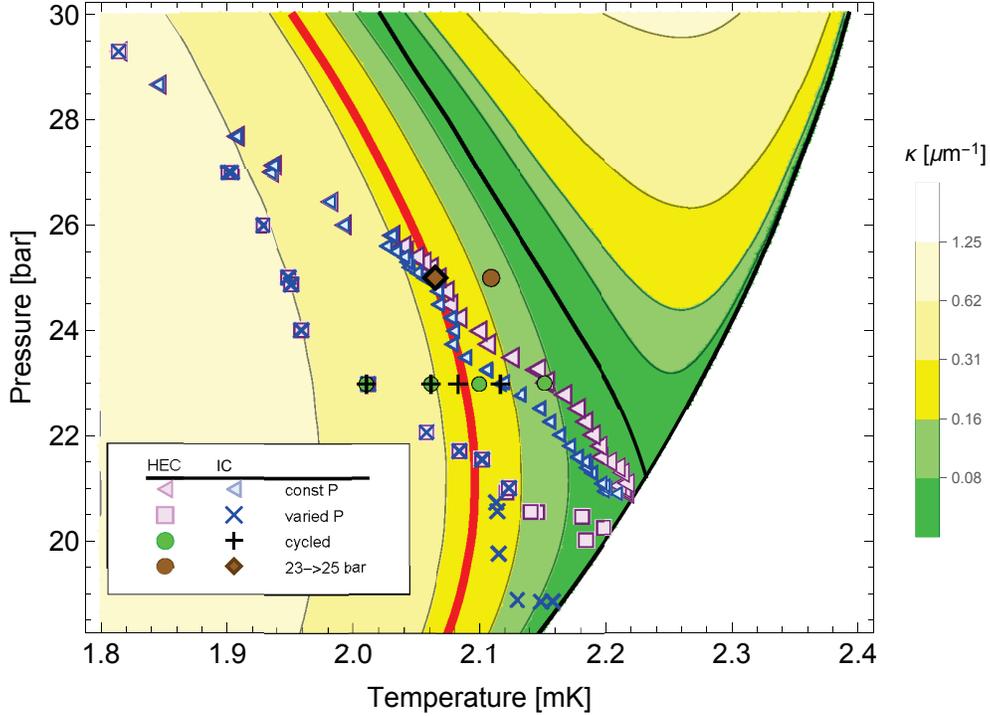} 
\caption{
  {\bf  Curvature of stable domain walls.}
  Contours show  $\kappa_A/\sin(\theta_A)$ and $\kappa_B$ in the A and B portions of the phase diagram.  Smallest contour (between the two shades of green) corresponds to 0.078 $\mu$m$^{-1}$, while each subsequent contours is a factor of 2 larger.
  $\kappa$ is the mean curvature of a stable A-B domain wall, and $\theta$ is the contact angle of a domain wall with a surface.  Here we assume minimal pairbreaking boundary conditions.  
  In the A phase, $\kappa_A/\sin(\theta_A)$ quantifies the inverse size of orifice which can block the motion of an A-B domain wall, while $\kappa_B$ represents the same quantity for the B phase.  
  Under the assumption that the B phase is seeded from chambers with small openings, the largest A phase value of $\kappa_A/\sin(\theta_A)$ will set the $\kappa_B$ where the A$\to$B transition is observed. Red line shows a contour, $\kappa=0.25~\mu$m$^{-1}$, which roughly corresponds to where domain walls pass freely through the 1.1 $\mu$m channel between the HEC and IC, corresponding to a contact angle of 74$^\circ$.  To the left of this line  transitions in the two chambers always occur simultaneously.   
 }
\label{fig::6_Zhang1}
\end{figure}

In Figure~\ref{fig::6_Zhang1} we show contours of constant $\kappa_{A}/\sin(\theta_{A})$ and constant $\kappa_{B}$, calculated using a Landau-Ginzburg theory and assuming minimal pairbreaking boundary conditions, corresponding to smooth surfaces. (See Supplementary Notes 2 and 3. Supplementary Note 5 deals with the results obtained for maximally pairbreaking boundary conditions.  There is some ambiguity in the temperature dependence of the Landau-Ginzburg parameters, and Supplementary Note 6 discusses an alternative model.)  As can be seen, $\kappa$ vanishes at the equilibrium A-B transition, where the two phases have the same free energy.  It also vanishes at $T_c$.  The dark green regions show where it is small.  The largest values of $\kappa_A/ \sin(\theta_A)$ are found at high pressure, and the largest values of $\kappa_B$ are found at low temperature.  Our model would predict  that for a given cooling trajectory $\kappa_B$ at the A-B catastrophe line will coincide with the largest value of $\kappa_A/\sin(\theta_A)$ encountered while cooling:  ie. $\Gamma=\kappa_B \sin(\theta_A)/\kappa_A =1$.  For example, a constant pressure cooled trajectory at 23.25 bar will almost touch the contour between the light and dark green regions in the A phase.  The A-B transition is therefore expected to be at the same contour in the B phase.  All of the varied-pressure trajectories pass through the A phase close to the contour that separates the two lightest shades of yellow -- and one therefore expects the catastrophe line to follow the corresponding B contour.

\begin{figure}
\centering
\includegraphics[width=\textwidth]{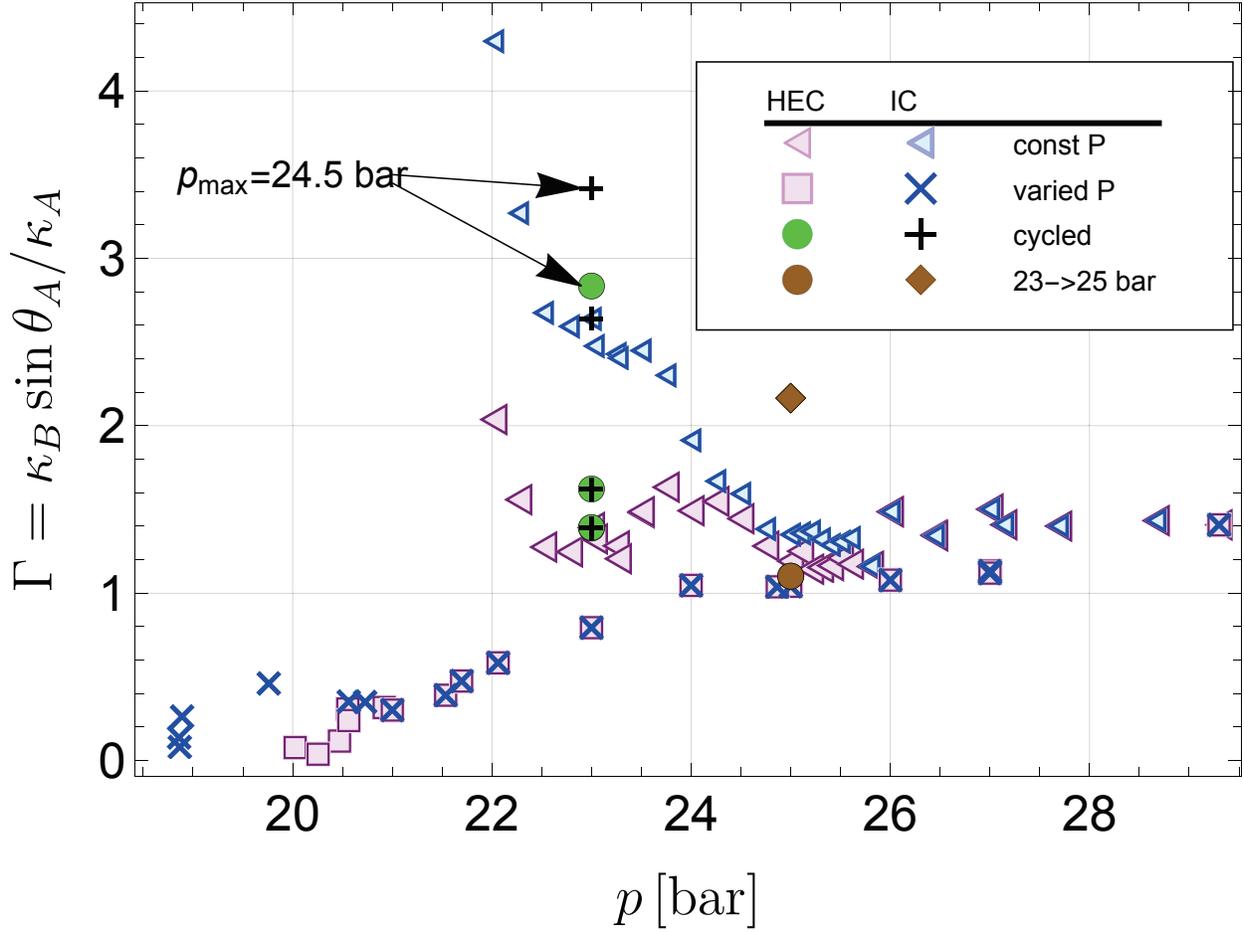}
\caption {
 {\bf Pressure dependence of $\Gamma=\kappa_A/\sin(\theta_A)$}.  Ordinate, $\Gamma$ is the ratio of the domain wall curvature at catastrophe point, $\kappa_B$, to the largest scaled curvature traversed in A phase, $\kappa_A/\sin(\theta_A)$. 
  The contact angle $\theta_A$ depends on the boundary condition: here we use minimal pairbreaking.  The abscissa shows the pressure at the observed A-B transition.
  The collapse of the data around unity above $p=24$ bar suggests that trapped pockets of B are responsible for the observed A-B transition at high pressure.}
 
\label{fig::7_Erich2}
\end{figure}

To better quantify the data, in Figure ~\ref{fig::7_Erich2} we plot the ratio $\Gamma=\kappa_B \sin(\theta_A)/\kappa_A $ {\it vs.} the pressure at which the A-B transition was observed. 
There is a remarkable data collapse for all pressures above 24 bar, despite the fact that the trajectories (pressure-varied or constant-pressure) are very different.  The ratio is 
larger than the expected value of 1, and is essentially constant.  Variations in the geometry or boundary conditions could cause this ratio to be different from unity -- for example, the contact angle could be slightly less than what is predicted by the theory. Previous experiments have directly tested aspects of our model of the AB phase boundary\cite{Bartkowiak2004}, including measuring equilibrium contact angles, surface tensions, and surface energies at low pressure.  A number of  theoretical works have also addressed the   issue\cite{Kaul1980,Schopohl1987,Thuneberg1991}.

Below 24 bar the fixed pressure HEC and IC A-B transition data separate.  Below 20.5 bar, a similar separation occurs in the pressure varied runs.  These features naturally correspond to when the channel connecting the HEC and IC can no longer support a domain wall.
This feature  is apparent in  Figure~\ref{fig::6_Zhang1}, where we draw a red line which corresponds to the contour with $\kappa=0.25~\mu$m$^{-1}$.  To the right of this line the transitions in the IC and HEC occur independently, while to the left they occur simultaneously.  
The IC data points which follow this red line correspond to events where the pre-existing B phase in the HEC propogates through the channel, and do not represent independent nucleation events. This includes the points below the PCP accessed by depressurization and then cooling at constant pressure.  
As argued in Supplementary Note 3, the B phase can propogate into the channel when  $\kappa=\cos(\theta)/W$, where $W=1.1~\mu$m is the height of the channel. Our inferred contact angle ($\theta\sim 75^\circ$) is larger than typical values predicted by the Landau-Ginzburg theory with minimal ($\theta\sim 30^\circ$)
or maximal ($\theta\sim 60^\circ$) pairbreaking boundary conditions (See Supplementary Figure 4).  This may be a feature of the glass and silicon surfaces in the channel, or it may point towards limitations in  the accuracy of our theoretical model. 

Between 22.5 and 23.8 bar, the constant pressure cooled transitions in the HEC continue to agree with our model, with $\Gamma\sim 1$ (pink triangles).  Over the same range the IC data clusters near $\Gamma\sim 2.5$ (blue triangles), and it is likely that the transition is completely independent of the HEC.  This clustering suggests that a similar model may apply there, but with different surface geometries and boundary conditions, or the involvement of different order parameter structures.  The HEC contains sintered silver, with clear candidates for B-containing chambers.  The IC incorporates as-machined coin silver surfaces, with  no obvious cavities which could be playing the role of the sinter in the HEC. 

An additional potential mechanism for heterogeneous nucleation involves the presence of surface defects, or features, which favour a distorted order parameter. The simplest model would treat this as a B phase seed, pinned at the surface with an associated A-B interface. The model of the catastrophe line 
would be analogous to the one we proposed for the sinter \cite{Balibar2000}. Given the multicomponent nature of the superfluid $^3$He order parameter (a complex 3 $\times$ 3 matrix), the nature of the spatially dependent order parameter of this seed region is complex. The path dependence could reflect 
evolution of the order parameter structures that alter the energetics of 
the transformation from the A phase  
to the B phase, 
without the benefit of an actual interface that would be present if a “seed” of B phase were present. 

The curvature $\kappa_A$ vanishes at the as one approaches the polycritical point from above, and hence  $\Gamma$ diverges near there for all of the constant pressure data.  At these pressures the distribution of B seeds are likely determined by kinetic processes occurring during the normal-superfluid transition rather than details of the cooling trajectory.  As emphasized in our previous work\cite{LotnykPRL2021}, it is surprising that we form the A phase when cooling at pressures below the PCP even in magnetic fields below 0.1 mT.

Below 24 bar, $\Gamma$ falls for the pressure varied data.  This suggests that a separate mechanism is at play:  The A-B transition occurs before our model would predict.  Below 20.5 bar the transition in the IC and HEC separate.  The ratio $\Gamma$ for the HEC data continues to fall, further indicating a mechanism in the HEC which goes beyond our model.  Between 20.5 and 19 bar the transition in the IC is likely not an independent nucleation event, but rather due to the  A-B domain wall breaking through the channel (corresponding to the red line in Figure~\ref{fig::6_Zhang1}).  This appears as a plateau Figure~\ref{fig::7_Erich2}.  The cluster of IC transitions at 19 bar are likely independent nucleation events.

Figure~\ref{fig::7_Erich2} contains two additional outliers.  The green discs and black crosses show $\Gamma$ for the pressure cycled transitions depicted in Figure~\ref{fig::5_QvsT}.  The trajectories that cycled to 26 bar and 27.5 bar agree very well with the rest of the data.  The trajectory that cycled to 24.5 bar, however, shows more supercooling than expected, and a surprisingly large value of $\Gamma$.  While we do not understand why the HEC shows such a large degree of supercooling, the IC transition coincides with the red line in Figure~\ref{fig::6_Zhang1}, and is likely due to the physics of the superfluid in the channel connecting the chambers.
Similarly, the brown  diamond corresponds to the trajectory in Figure~\ref{fig::3_ConstantPressure}(b) which was cooled at p= 23 bar to 2.2 mK, pressurized to 25 bar, and then further cooled.  It also lies on the red line in Figure~\ref{fig::6_Zhang1} and is presumed to correspond to the B phase being conveyed through the channel.  The transition in the HEC for this trajectory (brown disc) agrees well with our model.

Finally we note that the domain wall between the A and B phases has a finite width, extending over a few temperature dependent coherence lengths (see Eqn. 5 in Supplementary Note 2 and Supplementary Figure 2). Near $T_c$ the temperature dependent coherence length diverges, $\xi(T)\approx \xi_{GL}(1-T/T_c)^{-1/2}$. The resulting ``thick" domain walls are  likely to have different elastic properties and may not have a  well defined curvature.  This feature may account for some of the decrease in the ratio $\Gamma$ plotted in Figure~\ref{fig::7_Erich2} for the pressure varied runs that extended to transitions near $T_c$.

While we have developed a coherent picture, we emphasize that
 a number of mysteries still remain.  
 First, we do not have an explanation for the appearance of the A phase upon crossing $T_c$ for a range of pressures below the tricritical point.  Second, we do not have a model for the nucleation of B seeds during the transition from the normal phase into the superfluid.  Third,  we have yet to establish the exact form of order parameter features which hold the those seeds.  This is particularly true in the IC, which lacks any natural cavities.


In conclusion, we find that the supercooling of the A phase can be extended considerably by transiting through high pressure regions where the A phase is more stable (measured by the ratio of the A-B free energy difference per coherence length to the surface tension). The path dependence observed here is remarkable, and is only possible because of the purity of $^3$He and the relatively large energy barriers between the superfluid phases.  The supercooled liquid is stable at pressures as low as 18.6 bar which can be contrasted to the lowest stable pressure for bulk $^3$He, 21.23 bar.  We found that the lifetime of the metastable fluid exceeded one day at 19.8 bar. For a significant part of the phase diagram the degree of supercooling  appears to be determined by the maximum value of $\kappa_A/\sin(\theta_A)$ encountered -- a quantity which corresponds to the inverse size of an aperture which can support an A-B domain wall.

Despite these insights, aspects of the A-B transition remains enigmatic. In the superfluid $^3$He  environment of this experiment we have made a study of the systematics of heterogeneous nucleation by exploring a variety of trajectories in the pressure-temperature plane. We have shown that the surface energy of the A-B interface (strongly dependent on p and T), and the contact angle with surfaces play a central role in this nucleation process. On the other hand in previous work on superfluid $^3$He confined in a nanofluidic cavity\cite{Zhelev17NC}  negligible supercooling was observed. 
The transformation from the A to the B phase involves a transit through a multi-dimensional landscape that could be hysteretic with pressure.
To develop the A-B phase transition as a model for the first order transitions in the early universe,
identification
of all mechanisms 
is essential. As we observe in our analog system, 
it is possible that the early universe was not homogeneous, but may have contained structures such as topological defects or primordial black holes which could play a role in 
the nucleation of first order phase transitions \cite{Hiscock:1995ma,Gregory:2013hja}.
Further studies will include those of superfluid $^3$He confined in nano-structured environments, in which nucleation is studied in precisely engineered volumes, coupled to bulk liquid through “valves” which effectively isolate that volume from nucleation events in the bulk liquid and heat exchanger, and in which NMR is used as a non-invasive probe. Similar structures might also be used to seed the non-equilibrium Polar phase and other phases that are not by themselves stable or naturally occurring in the bulk. In turn, analogs of these structures might provide insights into the underpinnings of transitions in the early Universe.

\section*{Methods}

\subsection{Fork operation.}

The two quartz forks were each driven at constant voltage (small enough so that no drive dependent heating was observed) from a signal generator. A current preamplifier was used as the first stage of amplification before the received signal was 
sent to a lock-in amplifier. The lock-in's  reference frequency was ported from a signal generator. By  measuring and fitting the (complex) frequency dependent non-resonant signal in the circuit, the (anti-symmetric) quadrature component of the received signal (after background subtraction) was used to infer the difference between the drive frequency and the resonant frequency, while the in-phase component was used to infer the ``$Q$" or Quality factor of the fork. The forks were maintained within 10 Hz of the resonant frequency (near 32 kHz) with $Q$ factors varying from $\approx$ 40 at $T_c$, to about 200 at the lowest temperatures at high pressure. In operation the forks could track the $Q$ well without attention. At the A-B transition, Figure~\ref{fig::5_QvsT}, the $Q$ increased by $\approx$ 10 abruptly, providing a clear signature of the transition. 

\subsection{Thermometry.}
The temperature of the HEC detected at $T_c$ was found to lag the temperature of the melting curve thermometer (mounted on the demagnetization stage) by only 1-2 $\mu$K providing the warming and cooling rates were less than 10 $\mu$K/hr. We estimate the accuracy of our inferred A-B transition temperatures to be $\pm$ 3 $\mu$K, as long as the cooling rate was held constant in a given A-B transition run. The cooling rate of the nuclear stage was controlled by adjusting the rate of decrease of the current in the magnet and could be reliably set to be a constant 10 $\mu$K or even held constant ($\pm$ 3 $\mu$ K) for periods as long as a day. The temperature of the fork in the IC lagged that of the HEC by $\approx$ 15 $\mu$K (inferred by observing the differences in the observed $T_c$ while cooling at constant temperature).  In all graphs the data has been adjusted for this lag. The inferred temperature of $^3$He in the IC were similarly adjusted appropriately while cooling at constant pressure if the supercooled transitions in the IC and HEC occurred at different times (ie below 24.5 bar see Figure~\ref{fig::3_ConstantPressure})(a). 

\subsection{Pressure control.}
Pressure was regulated using a temperature controlled ``bomb" consisting of a $\approx$ 10 cm$^3$ volume in the form of a 9.5 mm diameter stainless steel tube.  An insulated Nichrome wire was wound on this tube and was connected to a 0-25 W power source whose output was set by a digital proportional integral and differential controller. The bomb was semi-isolated from the lab environment by being contained in a large cylindrical tube 5 cm id x 25 cm long open at both ends and mounted vertically.  The pressure could be monitored by a digital Heise DXD 0-40 bar pressure gauge, allowing for high resolution with minimum volume in the system. A 0.3 cm$^3$ volume filled with silver sinter was used as an additional heat sink to thermalize the $^3$He before it entered into the main HEC chamber. In this way we were able to vary the pressure by as much as 5 bar/day without incurring significant heating, allowing large pressure changes to be effected quote rapidly.  A-B transitions were observed while the pressure was constant or being varied by 0.7-1.3 bar/day in order to avoid any issues with viscous heating in the channel.
Due to the limited heat we could apply to the bomb, pressure changes greater than 5 bar required multiple steps, where we would stop and reset the pressure in the external gas handling system.

\begin{addendum}
 \item This work was supported at Cornell by the NSF under DMR-2002692 (Parpia), PHY-2110250 (Mueller), and in the UK by EPSRC under EP/J022004/1 and by STFC under ST/T00682X/1. In addition, the research leading to these results has received funding from the European Union’s Horizon 2020 Research and Innovation Programme, under Grant Agreement no 824109. Fabrication was carried out at the Cornell Nanoscale Science and Technology Facility (CNF) with assistance and advice from technical staff. The CNF is a member of the National Nanotechnology Coordinated Infrastructure (NNCI), which is supported by the National Science Foundation (Grant NNCI-1542081).

 \item[Correspondence] Correspondence and requests for materials should be addressed to J.M.P. \newline (email: jmp9@cornell.edu).

 \item[Author contributions] Experimental work and analysis was principally carried out by Y.T. with early contributions by D.L and A.E. assisted by A.C. with further support from E.N.S. and J.M.P.  Presentation of figures was the joint work of Y.T and A.E. assisted by K.Z., M.H. and D.L.. N.Z had established most of the routines for the phase locked loop operation of the quartz fork for earlier experiments. E.M significantly contributed to exploration of the phase diagram and the writing of the manuscript, and N.Z and T.S.A. established and carried out the nano-fabrication of the channel. M.H. and K.Z. in conjunction with E.M. explored the relationship of $\kappa$ to $\sigma$ and calculated the contours of constant $\kappa$. J.M.P. supervised the work and J.M.P., E.M. and J.S. had leading roles in formulating the research and writing this paper. All authors contributed to revisions to the paper.

 \item[Data Availability] The data that supports this study will be made available through Cornell University e-commons data repository at https://doi.org/10.7298/4fhq-e356. 

\item[Competing Interests]
The authors declare that they have no competing interests.
\end{addendum}

\section*{References:}

\clearpage
\pagebreak

\makeatletter 

\renewcommand \thesection {Supplementary Note \@arabic\c@section:}

\makeatother

\setcounter{figure}{0}
\renewcommand*{\citenumfont}[1]{S#1}
\renewcommand*{\bibnumfmt}[1]{[S#1]}

\section{Experimental details}

\begin{figure}[H]
 \renewcommand{\figurename}{Supplementary Figure}
\centering
\includegraphics[width=0.7\linewidth, keepaspectratio]{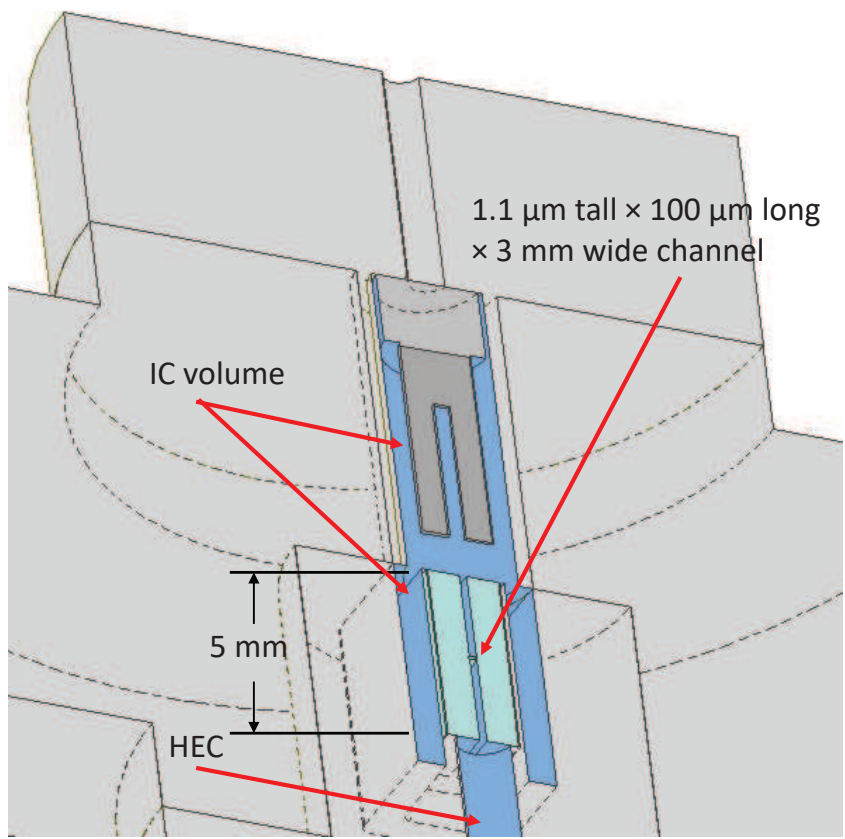}
\caption{Schematic of the IC. The assembly comprising the silicon and glass channel is shown as light blue and is 5 mm in length. The channel separating the IC (top) and HEC (region below the channel) is in the form of a wide septum with a letterbox shape 100 $\mu$m long $\times$ 1.1 $\mu$m tall $\times$ 3 mm wide, and is shown in cross-section with the channel width not visible here.  Most of the open volume resides in the bulk liquid surrounding the channel holder and around the quartz fork located above the channel. There are regions where close fitting coin-silver components that comprise the cell structure are separated by gaps of order 25 $\mu$m that may promote the A phase. All surfaces of the coin-silver components are as-machined metal.}
\label{fig::5_IC}
\end{figure}

The measurement of the AB transition occurs independently (at low pressures in pressure varied transitions and below $\sim$ 24 bar under constant pressure cooling) in the small volume isolated chamber (IC). The IC is connected to a chamber containing a sintered silver heat exchanger (HEC) through a 1.1~$\mu$m high, 3~mm wide, and 100~$\mu$m long channel with 200~$\mu$m tall $\times$ 3~mm wide $\times$ 2.45~mm long ``lead-in" sections at either end. The 1.1 $\mu$m height section appears to be filled with the B phase at all temperatures and pressures to the left of the contour highlighted in Fig.~5 of the main paper.  To the right of this line the channel isolates the two chambers, and the transitions occur independently.  To the left of this line, the B phase nucleated in one chamber will grow into the other. The physical arrangement is illustrated in Fig. 2 of the main paper. The channel (Supplementary Figure~\ref{fig::5_IC}, shown in light blue) was nanofabricated in 1~mm thick silicon, capped with 1 mm thick sodium doped glass, anodically bonded to the silicon\cite{Zhelev18RSI} and then glued into a coin silver carrier. 

The volume of the IC is estimated to be 0.14 $\pm$ 0.02 cm$^3$, and the area of all wetted surfaces in the IC was estimated to be  14.5 $\pm$ 0.5 cm$^2$. In comparison, the HEC had a volume of 0.72 $\pm$ 0.1 cm$^3$ and
a surface area of 3.5 $\pm$ 0.5 m$^2$ due to the heat exchanger.

\pagebreak
\section{Landau Ginzburg theory}
Superfluid $^3$He has a complex order parameter which can be represented by a 3 $\times$ 3 matrix $A_{\alpha j}$, where $\alpha$ corresponds to a spin angular momentum index, and $j$ an orbital angular momentum index: both take on 3 possible values: $x,y,z$.  The $2\times 2$ gap matrix is $\Delta =\sum_{\alpha j}  i\sigma_\alpha \sigma_y A_{\alpha j} p_j$, where the momentum $\bf p$ can be taken to lie on the Fermi surface.  
The $A$-phase corresponds to a rotation of $A_{\alpha j}= f_\alpha (\delta_{jx}+i\delta_{jy})$, for which the orbital angular momentum points in the $\hat z$ direction, and the gap matrix has a node in that direction.  If one ignores dipolar interactions, $f$ is a completely arbitrary vector of fixed magnitude.  The $B$ phase corresponds to a rotation of $A_{\alpha j}\propto \delta_{\alpha j}$.  This is an isotropic fully gapped state with the spin and orbital angular momenta entangled with one-another.

In equilibrium the order parameter minimizes a free energy $\Omega=\int (f_{\rm bulk}+f_{\rm grad} ) d^3 r$.  We neglect dipolar terms, which are on a smaller energy scale. Near $T_c$, the resulting bulk free energy density is written\cite{Vollhardt2003,wiman1,wiman2016,ChoiPRB07}
\begin{eqnarray}
f_{\rm bulk} &=&
\alpha {\rm Tr}(A A^\dag)
+\beta_1 |{\rm Tr} (A A^T)|^2
+\beta_2 |{\rm Tr} (A A^\dag)|^2\nonumber\\&&
+\beta_3 {\rm Tr} (A A^T)(A A^T)^*
+\beta_4 {\rm Tr} (A A^\dag)^2
\nonumber
+\beta_5 {\rm Tr} (A A^\dag )(A A^\dag)^*.
\end{eqnarray}
The coefficients are 
\begin{eqnarray}
\alpha&=& \frac{1}{3} N(0)\left(\frac{T}{T_c}-1\right)\\
\beta_j&=& \beta_0 \tilde \beta_j\\
\beta_0 &=& -\frac{N(0)}{(\pi k_B T_c)^2}\frac{7}{240} \zeta(3)
\end{eqnarray}
We use transition temperatures $T_c(p)$ tabulated by Greywall\cite{wiman1,Greywall86SH}, converted to the PLTS scale using the procedure in Ref. \citenum{PLTS}.  The density of states at the Fermi Surface is $N(0)$, and $T_c$ is the normal-superfluid transition temperature.

At $T_c$, the strong coupling $\beta$-coefficients are $\tilde \beta_j=\tilde \beta_j^{\rm WC}(p) +\Delta\beta_j(p)$, and we use the $\Delta\beta_j(p)$'s which are tabulated by Regan, Wiman, and Sauls (RWS)\cite{regan}, who calculated the values from first principles.  The weak coupling $\beta$'s are
$\tilde\beta_1^{\rm WC}=1,\tilde\beta_2^{\rm WC}=
\tilde\beta_3^{\rm WC}=\tilde\beta_4^{\rm WC}=-2,
\tilde\beta_5^{\rm WC}=2$.  
We rescale the RWS pressures by a factor of 1.01 so that the polycritical point occurs at the experimental $p_{PCP}=21.22$ bar.

We estimate the temperature dependence of the strong-coupling coefficients by taking $\tilde \beta_j^{SC}=\tilde \beta_j^{\rm WC}(p) +(T/T_c)^{\eta(p)} \Delta\beta_j(p)$.  We choose $\eta(p)$ in order to correctly match the experimental A-B transition temperature\cite{Greywall86SH}, which are well approximated by $T_{AB}/T_c(p)\approx 1 - 0.174 (p- p_{PCP})/{\rm bar}$.  As explained below,  the A-B equilibrium line is at $\beta_1+\beta_3/3-2 \beta_5/3=0$.
We therefore take
\begin{equation}
\eta(p)=-\frac{\log -(\Delta \beta_1+\Delta \beta_3/3- 2 \Delta\beta_5/3)/
(\beta_1^{\rm WC}+ \beta_3^{\rm WC}/3- 2 \beta^{\rm WC}_5/3)}
{\log(1 - 0.174 (p- p_{PCP})/{\rm bar})}.
\end{equation}
This results in $\eta=1.79,1.58,1.26$ at $p=19,24$, and $29$ bar.
Previous works\cite{wiman1,regan} which used $\eta(p)=1$, find a somewhat shifted A-B line.  Some of the quantities we calculate are quite sensitive to the location of the transition, necessitating the extra empirical parameter.

We have also explored other approaches to correct for this shift in the A-B line.  Supplementary Note 6 presents one of those, where we simply rescale the parameters, $\tilde \beta_j^{SC}=\tilde \beta_j^{\rm WC}(p) +\alpha(p)  \Delta\beta_j(p) T/T_c$. 
We have also repeated the analysis with alternative strong-coupling parameters in the literature \cite{ChoiPRB07,wiman1}.  All models give similar predictions, once the location of the equilibrium AB transition line is corrected.

The gradient terms in the free energy are
typically written as
\begin{equation}
f_{\rm  grad}=
\sum_{\mu j k} K_1 |\nabla_k A_{\mu j}|^2
+ K_2  \nabla_j A_{\mu j}^* \nabla_k A_{\mu k}+K_3 \nabla_j A_{\mu k}^* \nabla_k A_{\mu j}
\label{eq:gradterm}
\end{equation}
with $K_1=K_2=K_3=K$, and
\begin{equation}\label{kval}
K=\frac{7 \zeta(3)}{60} N(0) \xi_0(p).
\end{equation}


It is convenient to scale the energy and order parameter in order to make the equations dimensionless.  Thus we take
$A=A_0 \tilde A$,
$f= |\alpha| A_0^2 \tilde f$, and choose $A_0$ such that $\beta_0 A_0^4 =\alpha A_0^2$.  We also rescale lengths by the temperature dependent correlation length $\xi= \xi_{GL}/\sqrt{1-T/T_c}$, to arrive at a dimensionless equation.    Here the Ginzburg-Landau coherence length is $\xi_{GL}=\sqrt{7\zeta(3)/20}\xi_0$.

For the bulk A and B phases we take $\tilde A_{\alpha j}=a \delta_{\alpha j}(\delta_{jx}+I\delta_{jy})/\sqrt{2}$ and
$\tilde A_{\alpha j} = a \delta_{\alpha j}/\sqrt{3}$.  These ansatzes yield free energy densities of the form $\tilde f=-a^2+a^4 \beta_\Phi$, where $\Phi = A,B$ denotes the phase, and the quartic coefficients for each phase are $\beta_A=\beta_{245}$ and $\beta_B=\beta_{12}+\beta_{345}/3$.  Here we use the convention that appending indices yields the sum,  so $\beta_{234}\equiv\beta_2+\beta_3+\beta_4$.
Minimizing with respect to $a$, gives $\tilde f=-1/(4 \beta)$.
The equilibrium AB transition occurs when these energies are equal to one-another, ie $\beta_1+\beta_3/3-2 \beta_5/3=0$.

\begin{figure}
\centering
 \renewcommand{\figurename}{Supplementary Figure}
\includegraphics[width=0.85\textwidth]{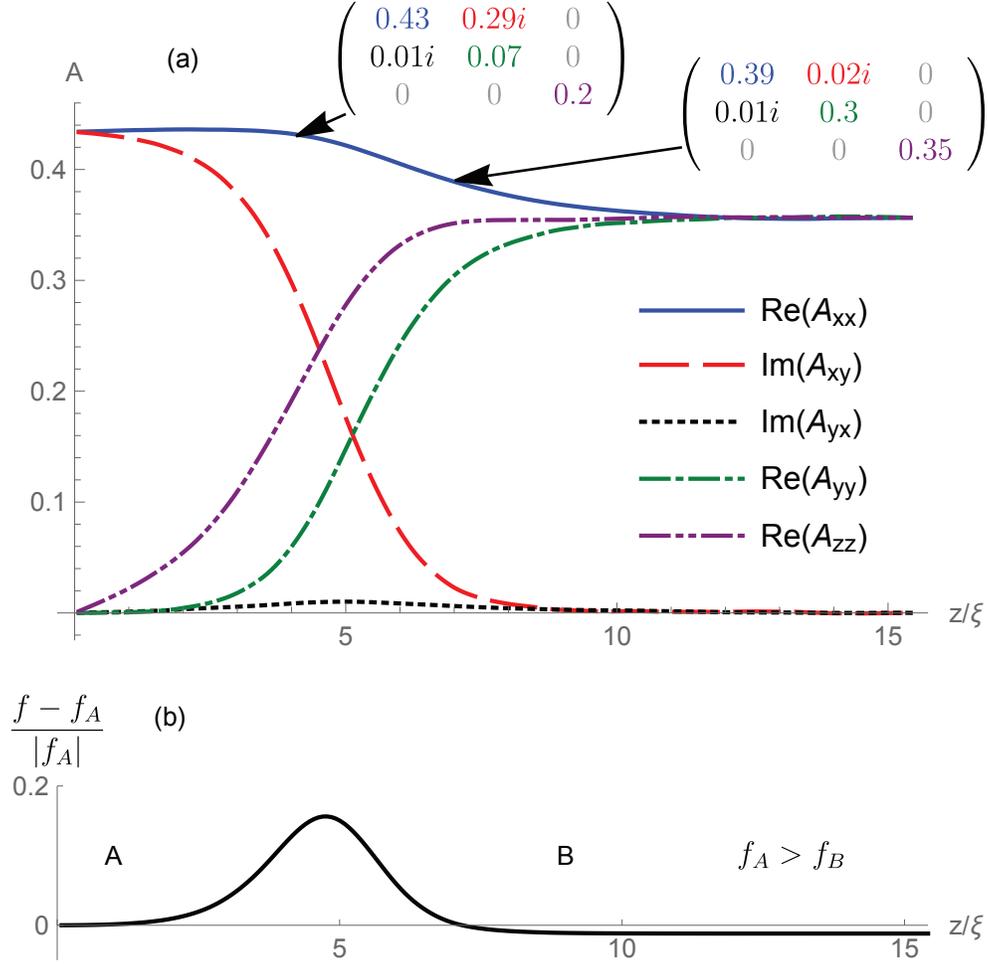}
\caption{\small
  {\bf Spatial dependence of the order parameter in a A-B domain Wall.} (a) Components of the order parameter $A_{\alpha j}$ in a domain wall between the A phase (left) and B phase (right).  The order  parameter  is constrained to its bulk A and  B values at the edges.   Magnitudes are scaled to dimensionless units, as  described after  Eq.~(\ref{kval}).  (b)  Local  free energy density, $f$, in this domain wall, as compared to the energy density of the A-phase.  Here $T=2.1$ mK and $p=24$ bar, and the temperature dependent Ginzburg-Landau coherence length is $\xi=39$ nm.  For these parameters, the free  energy of the B-phase is somewhat lower than the A-phase.  The surface tension $\sigma$ is calculated  by integrating the  the curve in (b), accounting  for the offset between $f_A$ and $f_B$.
  }
\label{fig::domainwall}  
\end{figure}

We use a finite difference approach to find the structure of the AB domain wall, taking $\tilde A$ to be a function of one spatial coordinate, $x$.  We minimize the free energy with the constraint that $\tilde A$ takes on its bulk A and bulk B form at $x=0$ and $x=L$.  For the minimization we use a Broyden, Fletcher, Goldfarb, Shanno quasi-Newton method, discretizing space into 200 segments, each of which is 5\% of the correlation length.
Supplementary Figure~\ref{fig::domainwall} illustrates how the components of $\tilde A$ vary in space, as well as the local energy density.  We define the position of the domain wall by the location of the peak in the local energy density, $x_p$.  The surface tension is then extracted as $\sigma= f-f_A x_p - f_B (L-x_p)$.  Here $f_A,f_B$ are the bulk free energies.

We similarly find  the surface energy associated with bulk A or B phase contacting the chamber's wall: $\tau_A,\tau_B$.  We consider two extreme boundary conditions: minimal and maximal pairbreaking.  
The former models specular scattering of quasiparticles from the wall \cite{ambegaokar}, and the latter corresponds to quasiparticle retroreflection \cite{saulssurface}.

For maximal pairbreaking, the order parameter simply vanishes at the wall.
For minimal pairbreaking the component of the order parameter normal to the surface vanishes,  $\sum_{j} A_{\alpha j} n_j=0$, 
where $n_j$ are the components of the normal vector to the wall.
The other components have zero slope, $\sum_j n_j \nabla_j  A^\perp_{\alpha i}$. Here $ A^\perp_{\alpha i}= A_{\alpha i}-n_i \sum_j (A_{\alpha j} n_j)$ is the component of $A$ which is perpendicular to the surface. 

For minimal pairbreaking one can align the A phase order parameter so that $\tau_A=0$.  While there is some temperature and pressure variation, $\tau_B\approx 0.8 \sigma$.  For maximal pairbreaking, $\tau_A\approx 2\sigma$ and $\tau_B\approx 2.5\sigma$.  The fact that $\tau_A<\tau_B$ for maximal pairbreaking can be traced to the contribution of the various gradient terms in  Eq.~(\ref{eq:gradterm}).
\pagebreak

\section{Stability of a domain wall across an orifice}
In our experiment we observe path dependent hysteresis in the A to B phase transition.  
The simplest model which could account for this observation involves positing  that the cell contains a large number of small chambers connected to the main fluid via small orifices.  
Leggett and Yip\cite{LeggettYip1990} described such chambers as ``lobster pots," and used them to explain related hysteretic phenomena.
The orifices have a distribution of sizes, and each of the chambers contain B-fluid.  The bulk fluid is A.  The stability of an AB domain wall across such an orifice was theoretically studied by Viljas and Thuneberg\cite{Viljas2003}.  As we review below, each AB boundary is only stable if the orifice size $W$ is smaller than some critical value $W<W_c$, where $W_c$ depends on pressure and temperature.  If $W>W_c$, and we are in the part of the phase diagram where the A-phase is stable,  the bulk A-phase will rush into the chamber, destroying the B-seed.  Conversely, in the B-region of the diagram, the B-phase will rush out, converting the bulk to B.  The same argument can be used to analyze the ability of the 1.1 $\mu$m letterbox channel to prevent the B-phase from propagating from the HEC to IC.  

In this model, the system's memory comes from the distribution of chambers which contain B, and the catastrophe line is set by the largest orifice of a B-chamber. 
We emphasize that this chamber model is merely intended to be an easily analyzed toy.  It is likely  that the true B-seeds involve frustrated A textures, whose modeling is complex.  These may be bulk textures, or they may involve boundaries. Similar  energetic considerations will be at play -- regardless of the microscopics.  The relevant scales are the free-energy difference between the two phases, the domain wall surface tension, and the size of the texture.

The chamber model seems particularly relevant for describing the sinter in the HEC.
Even there, however, the geometry is likely more complicated.  As argued by Leggett and Yip\cite{LeggettYip1990}, the openings of the chambers can be tapered, or have some other asymmetry, which will lead to quantitative discrepancies.

We treat the A-B domain wall as an elastic membrane.  
Consequently, the equilibrium domain wall will be bowed, with the phase with higher free energy on the inside.
Balancing forces  implies that in equilibrium its mean curvature is $\kappa=\delta f/(2\sigma)$, where $\delta f$ is the difference in free energy density between the fluids on each side of the wall, and $\sigma$ is the surface tension\cite{soap}.  The mean curvature is the arithmetic mean of the two principle curvatures: spherical and cylindrical surfaces of radius $R$ respectively have $\kappa=1/R$ and $\kappa=1/(2R)$.  

\begin{figure}
\centering
 \renewcommand{\figurename}{Supplementary Figure}
(a)
\includegraphics[width=0.25\textwidth]{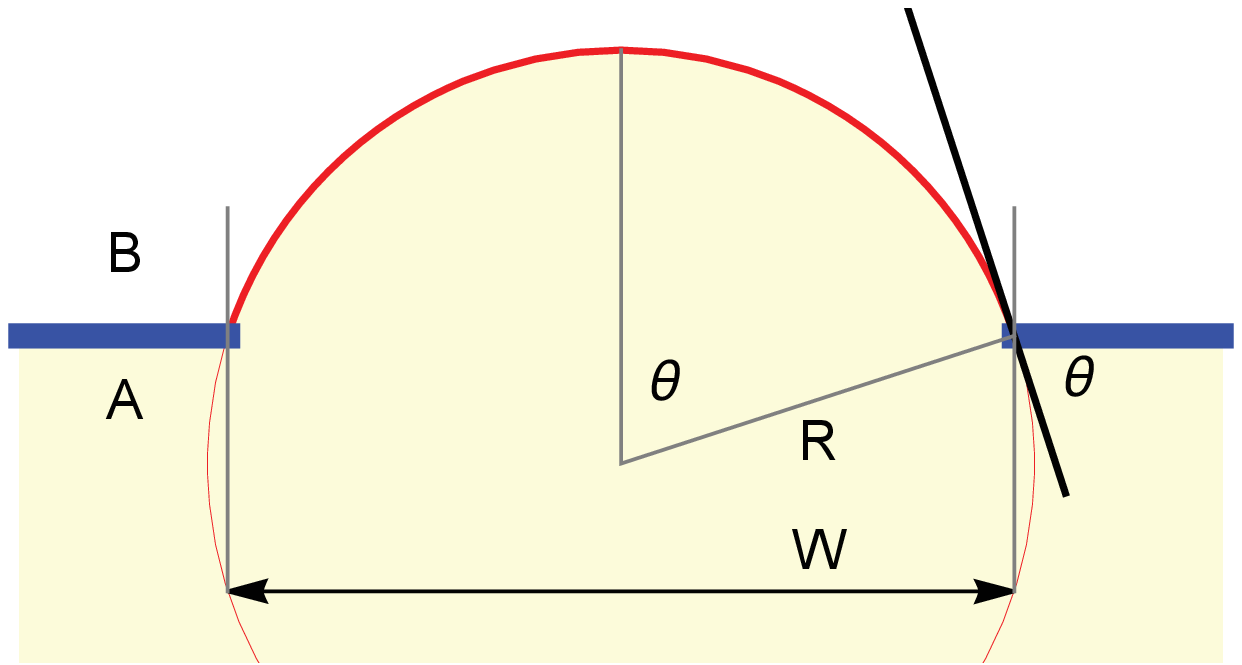}
(b)
\includegraphics[width=0.25\textwidth]{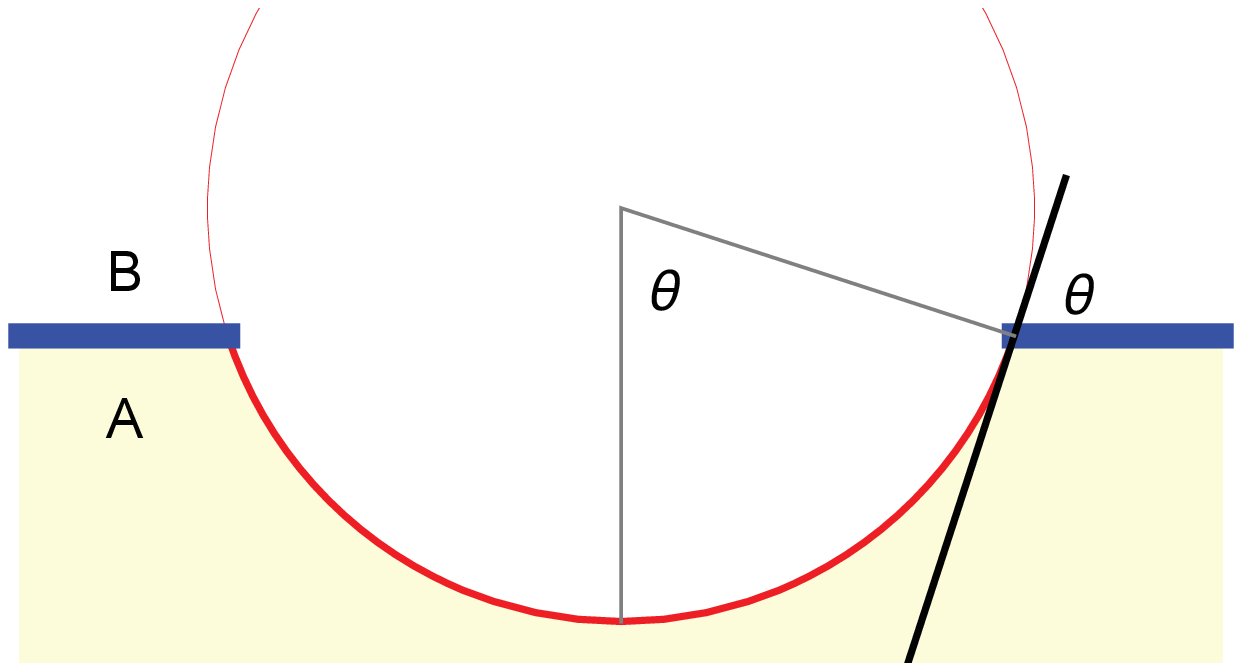}
(c)
\includegraphics[width=0.25\textwidth]{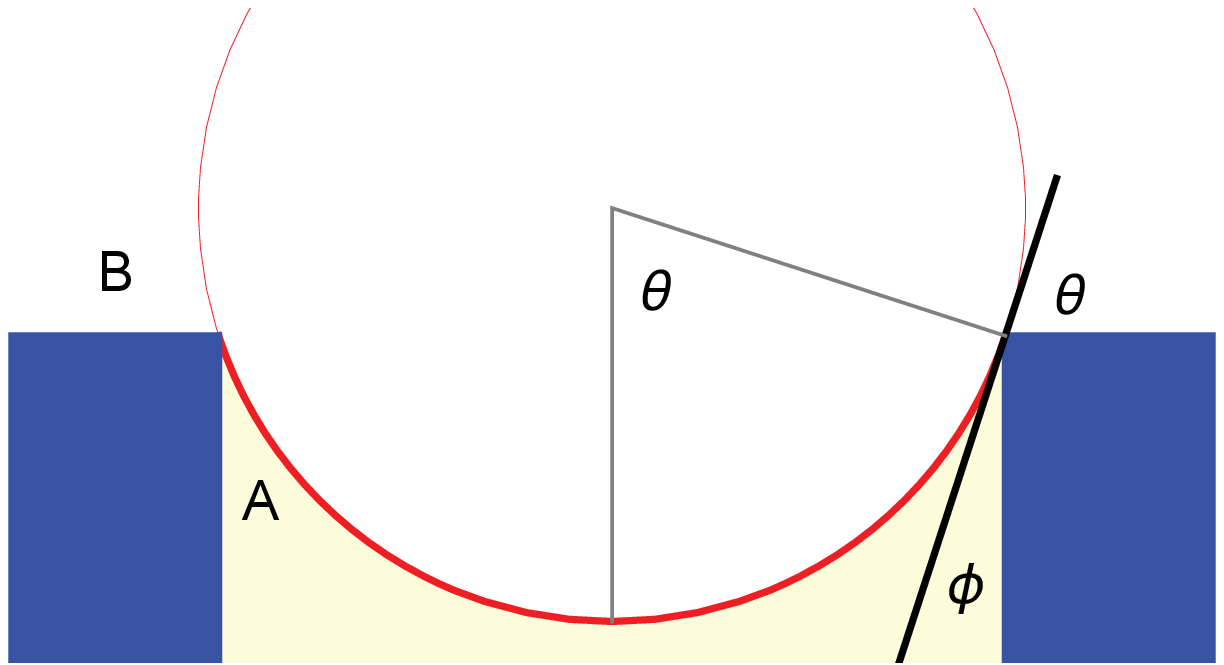}
\caption{\small
  {\bf Geometry of AB phase boundary.} Within the elastic model of the AB boundary, an equilibrium domain wall forms a surface of constant mean curvature.  Three geometries are depicted:  (a) and (b) show the case where the domain wall is pinned by a circular hole or a long slit.  In (a) the free energy density of the A phase is less than that of the B phase, while (b) shows the opposite.  Panel (c) depicts the case where the A phase is contained in a narrow channel.  
  }
\label{fig::dwimages}  
\end{figure}

Supplementary Figure~\ref{fig::dwimages} shows a number of simple geometries for a domain wall spanning a constriction.  Panels (a) and (b) show the case where the constriction can be modeled as a hole in a flat sheet.  In (a) the A phase has lower free energy than B, and the domain wall bows into the B region.  Equating forces at the edge  of the aperture  requires $\tau_B-\tau_A\leq 
\sigma\cos(\theta)$, where $\tau_B$ and $\tau_A$ are the surface energy associated with each phase, and $\theta$ is the angle that the surface makes with the walls -- which also corresponds to half the angle subtended by the opening, as measured from the center of the arc which defines the domain wall.  Consequently, the domain wall is stable if $W\leq 2R \sin(\theta)$.  Panel (b) shows the case where B has a lower free energy than A.  Since $\tau_A<\tau_B$, surface energies are never relevant and the domain wall is stable if $W<2R$.

Panel (c) of Supplementary Figure~\ref{fig::dwimages} shows the case where the constriction has parallel sides.  This models the letterbox channel connecting the HEC and IC.  Because $\tau_A<\tau_B$, such a configuration is never stable with B inside the chamber and A outside.  The illustrated configuration is stable as long as $\tau_B-\tau_A\geq\sigma\cos(\phi)$.  Here the angle that the domain wall makes with the surface, $\phi$, is complementary to the angle subtended by the domain wall, $\theta$.  Thus the domain wall is stable if $W\leq 2R (\tau_B-\tau_A)/\sigma$.  Specializing to the case of a slit, where $R=1/(2\kappa)=\sigma/\delta_f$, this expression simplifies to $W\leq 2 \delta \tau/\delta f$, which does not depend on the AB surface tension.  Equality coincides with the equilibrium A to B phase transition in a thin channel: $W \delta f = 2\delta \tau$.

The pressure and temperature dependence of the contact angle is illustrated in Supplementary Figure~\ref{fig::contact} for both minimal and maximal pairbreaking boundary conditions.  The A-phase surface energy vanishes for minimal pairbreaking, resulting in a significantly smaller contact angle.  Due to the behavior of the strong-coupling parameters, the contact angle is slightly larger at high temperatures and high pressures.

\begin{figure}
\centering
 \renewcommand{\figurename}{Supplementary Figure}
\includegraphics[width=0.6\textwidth]{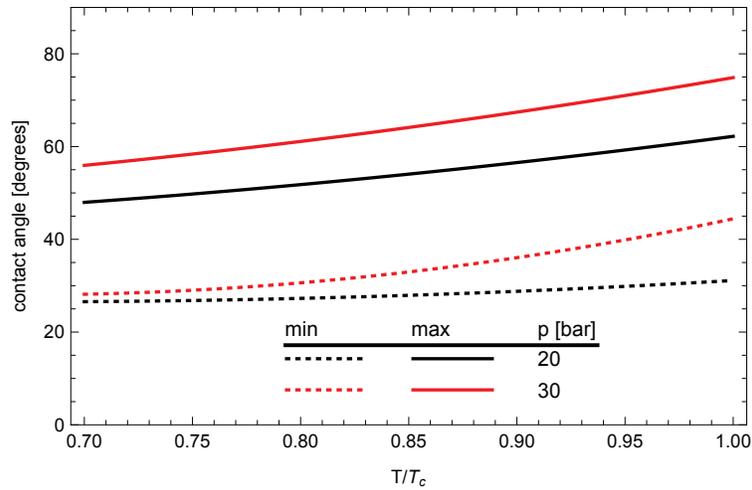}
\caption{\small
  {\bf Contact angle of an AB domain wall with a surface.} Solid lines: Maximal pairbreaking.  Dotted lines: Minimal pairbreaking.  Black: $p=20$ bar, Red: $p=30$ bar.  
  }
\label{fig::contact}  
\end{figure}

Previous experiments have directly tested
aspects of this elastic model of the AB phase boundary\cite{Bartkowiak2002,Bartkowiak2003,Bartkowiak2004}, including measuring equilibrium contact angles, surface tensions, and surface energies.  A number of  theoretical works have also addressed the   issue\cite{Kaul1980,Schopohl1987,Thuneberg1991}.

\pagebreak

\section{Comparing constant pressure and pressure varied A-B transitions}

\begin{figure}
 \renewcommand{\figurename}{Supplementary Figure}
\centering
\includegraphics[width=0.3\textwidth]{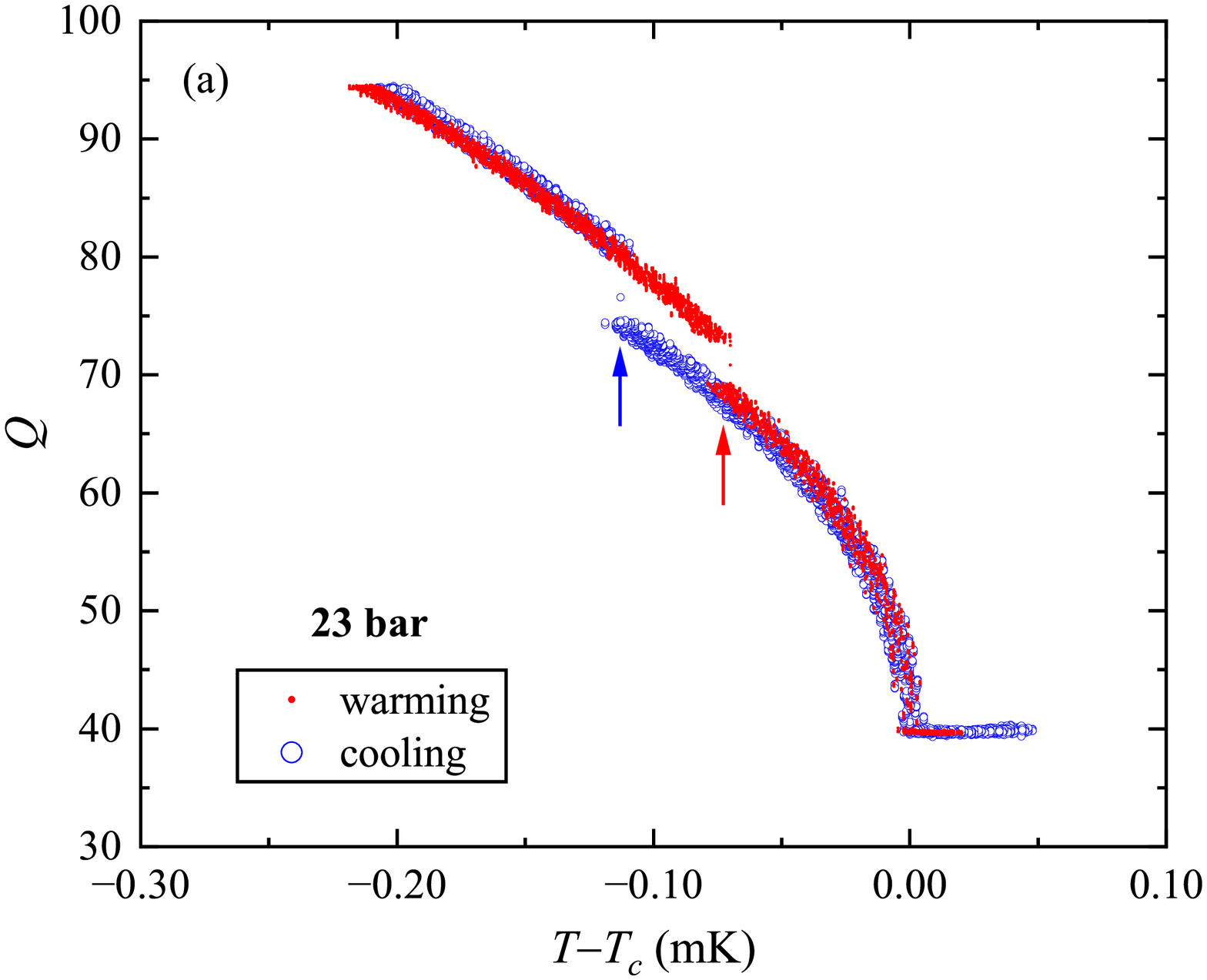}
\includegraphics[width=0.3\textwidth]{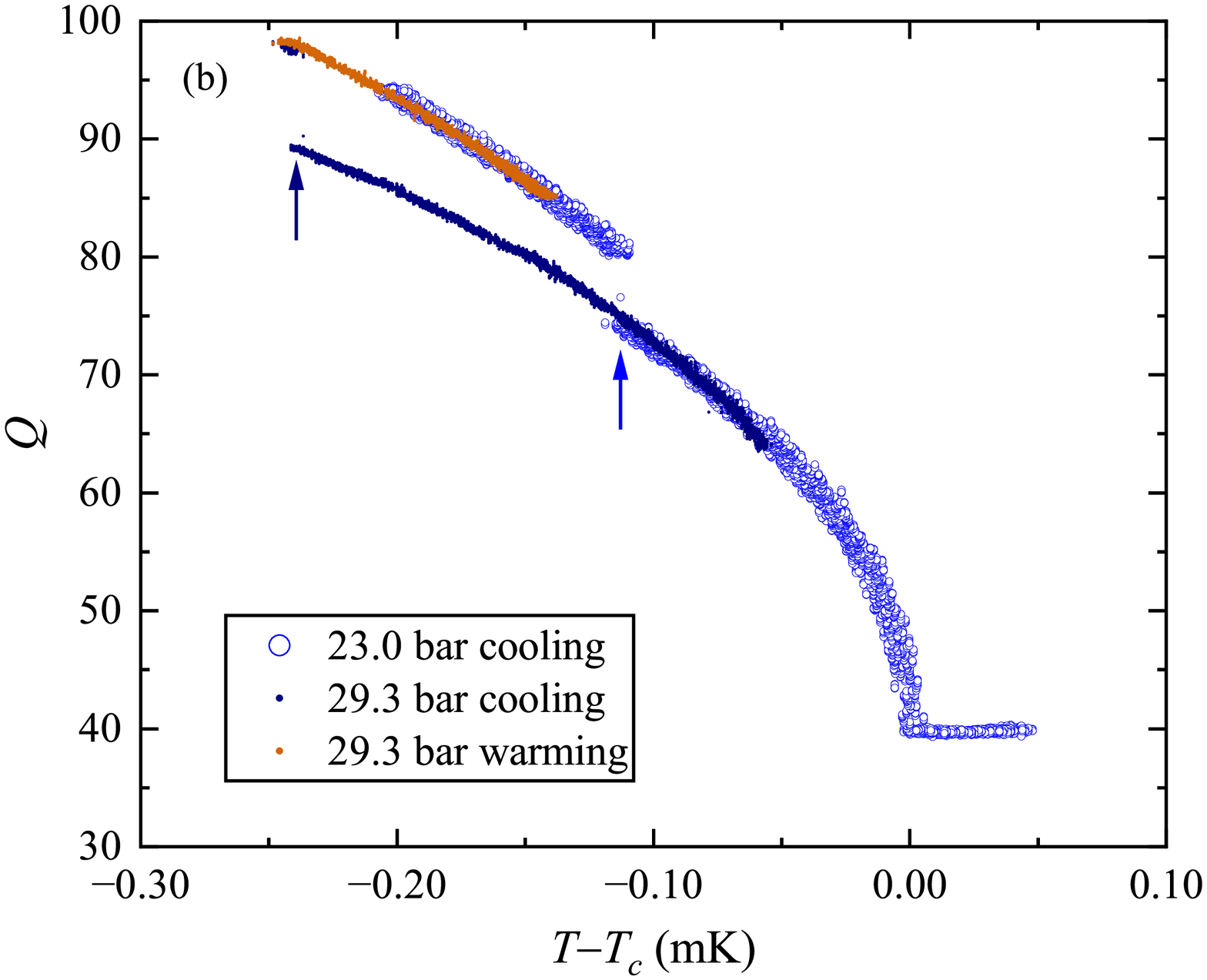}
\caption{\small
  {\bf $Q$ vs T in HEC.} (a) $Q$ {\it vs} $T$ of the HEC fork (blue open circles-constant pressure cooled at 23 bar) (red dots-warming). (b) Black points - depressurized at constant $Q$ from 29.3 bar to 23 bar, then cooled at constant pressure - extend the supercooling beyond that accessible at constant pressure (compare Blue and Black arrows). Only the part of the trajectory where $p=23$ bar is shown.  Orange points show part of the data obtained while warming in the B phase after depressurization. 
  }
\label{fig::4_Depress}  
\end{figure}

In the main document we established that the degree of metastability of the A-phase depended on the trajectory taken through the 
phase diagram.  Here we compare two trajectories.  Supplementary Figure~\ref{fig::4_Depress}(a) shows the quality factor of the HEC fork during cooling (blue) and heating (red) at 23 bar. Supplementary Figure~\ref{fig::4_Depress}(b) shows the same cooling data, and overlays the $Q$ measured while first cooled at 29.3 bar to 2.15 mK, and then depressurized to 23 bar (in the A phase) followed by cooling at constant pressure.  Data is only shown for the part of the trajectory at 23 bar.  In all cases the A-B transition is visible as a discontinuity in $Q$.  The degree of metastablity is enhanced by transiting through the 29.3 bar A-phase; however, the $Q$ in both phases is consistent, showing that the additional supercooling cannot be due to a pressure lag, heating effects, or a significant change in the properties of the bulk fluid.

\pagebreak

\section{Comparison of boundary conditions}

To highlight the role of boundary conditions, we reproduce the analysis used for Figs. 5 and 6 of the main paper, but using maximal pairbreaking boundary conditions.  As seen in Supplementary Figure~\ref{fig::maxpd}, the maximal pairbreaking contours of $\kappa/\sin \theta$ are shifted to slightly higher pressures compared to the minimal pairbreaking shown in Fig. 5 of the main paper.  This results in a larger value of $\Gamma$, plotted in Supplementary Figure~\ref{fig::maxratio}.  Given that our model predicts $\Gamma=1$, we conclude that minimal pairbreaking better describes the surfaces of the silver sinter.  This is consistent with previous experiments \cite{heikkinen2019}.

\begin{figure}
\centering
 \renewcommand{\figurename}{Supplementary Figure}
\includegraphics[width=0.5\textwidth]{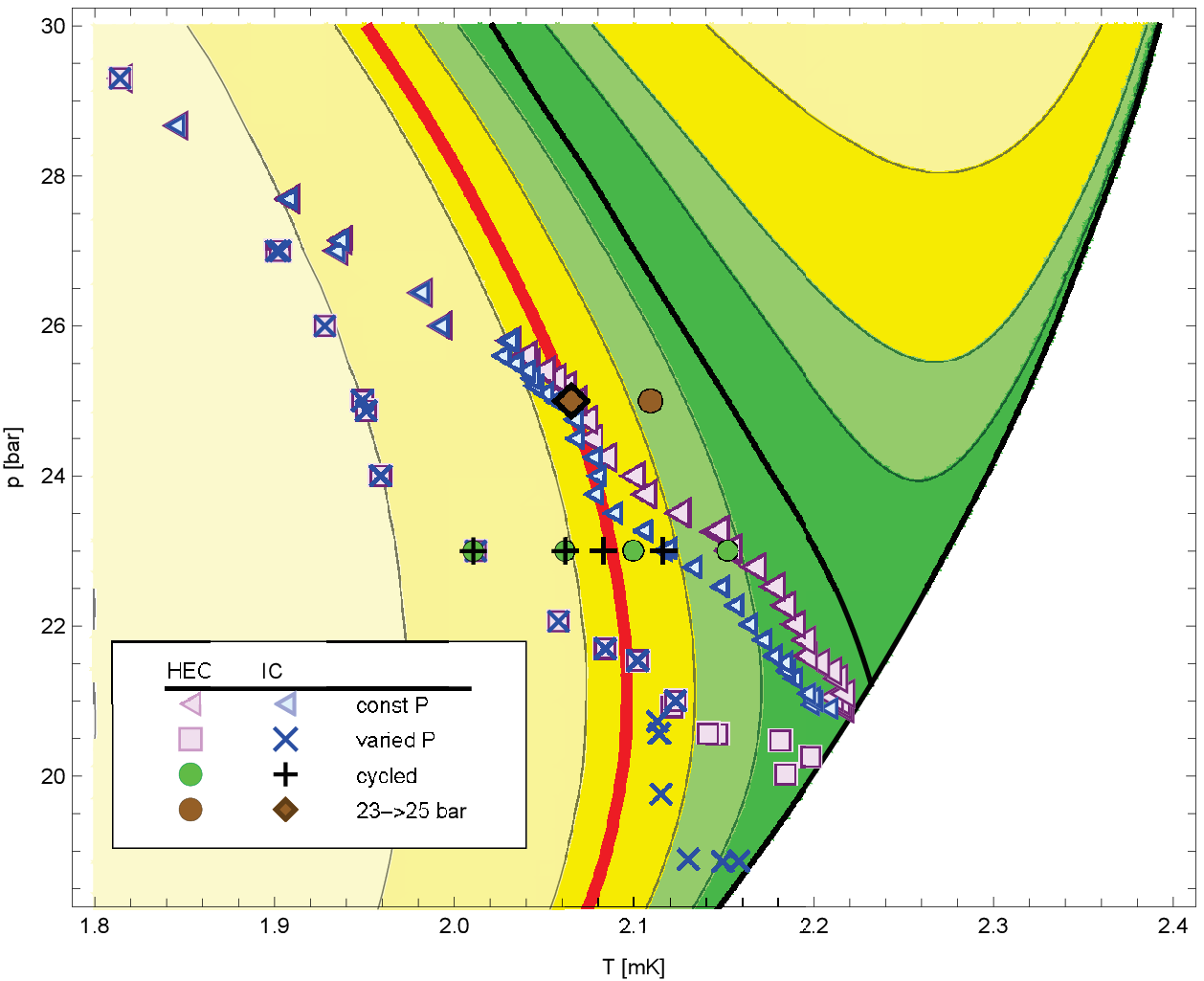}
\caption{
  {\bf  Curvature of stable domain walls, using maximum pairbreaking.}
  Contours show  $\kappa_A/\sin(\theta_A)$ and $\kappa_B$ in the A and B portions of the phase diagram, using maximal pairbreaking boundary conditions to calculate $\sin(\theta_A)$.  All values coincide with those of Fig 5 of the main text.  Here, because of the change in boundary conditions, the contours in the A-phase are shifted to slightly higher pressures.
 }
\label{fig::maxpd}
\end{figure}

\begin{figure}
\centering
 \renewcommand{\figurename}{Supplementary Figure}
\includegraphics[width=0.7\textwidth]{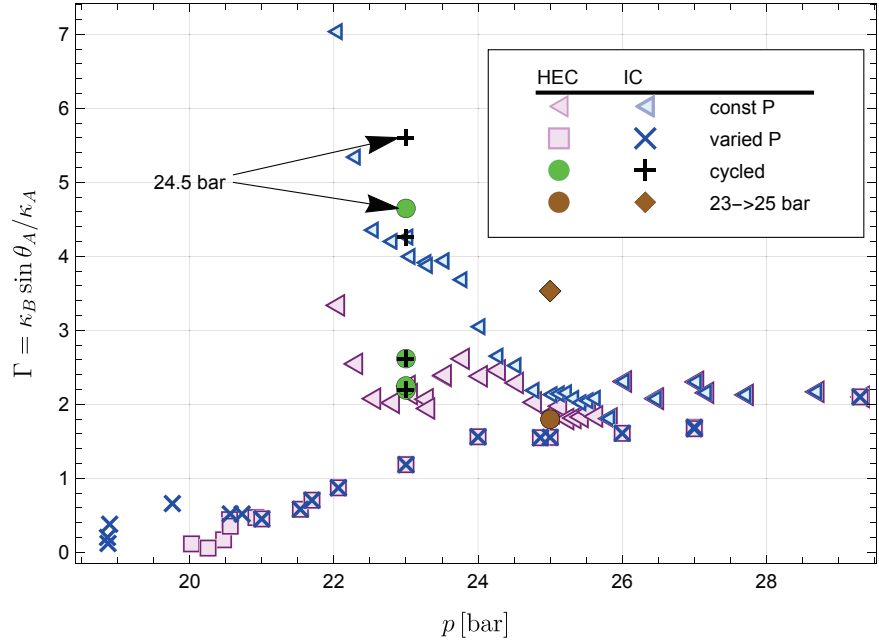}
\caption 
 {{\bf Pressure dependence of $\Gamma$ at catastrophe point for maximal pairbreaking boundary conditions.}  Ordinate, $\Gamma$ is the ratio of the domain wall curvature at catastrophe point, $\kappa_B$, to the largest scaled curvature traversed in A-phase, $\kappa_A/\sin(\theta_A)$. 
  The contact angle $\theta_A$ depends on the boundary condition: here we use maximal pairbreaking.  The abscissa shows the pressure at the observed A-B transition.  This graph should be compared to Fig. 6 of the main text, which used minimal pairbreaking boundaries.}
\label{fig::maxratio}
\end{figure}

\pagebreak

\section{Alternate renormalization of the Ginzburg-Landau parameters}\label{sec::lin}
In Supplementary Note 2, we explain how we introduce a  phenomenological scaling function $\eta(p)$ which allows the Landau-Ginzburg theory to correctly model the pressure dependence of the AB phase transition line.  Here we present an alternative approach, where we simply rescale the Regan, Wiman, and Sauls coefficients\cite{regan}, taking $\tilde \beta_j^{SC}=\tilde \beta_j^{\rm WC}(p) +\alpha(p)  \Delta\beta_j(p) T/T_c$.  

We again use the approximation that 
$T_{AB}/T_c(p)\approx 1 - 0.174 (p- p_{PCP})/{\rm bar}$.  Using the fact that  the A-B transition occurs when $\beta_1+\beta_3/3-2 \beta_5/3=0$,
We conclude
\begin{equation}
\alpha(p)=-\frac{\beta_1^{\rm WC}+\beta_3^{\rm WC}/3- 2 \beta^{\rm WC}_5/3}{\Delta \beta_1+\Delta \beta_3/3- 2 \Delta\beta_5/3} \frac{1}{1 - 0.174 (p- p_{PCP})/{\rm bar}}
\end{equation}
which yields $\alpha=1.03,0.98,0.96$ at $p=19,24,29$ bar.

\begin{figure}
\centering
 \renewcommand{\figurename}{Supplementary Figure}
\includegraphics[width=0.60\textwidth]{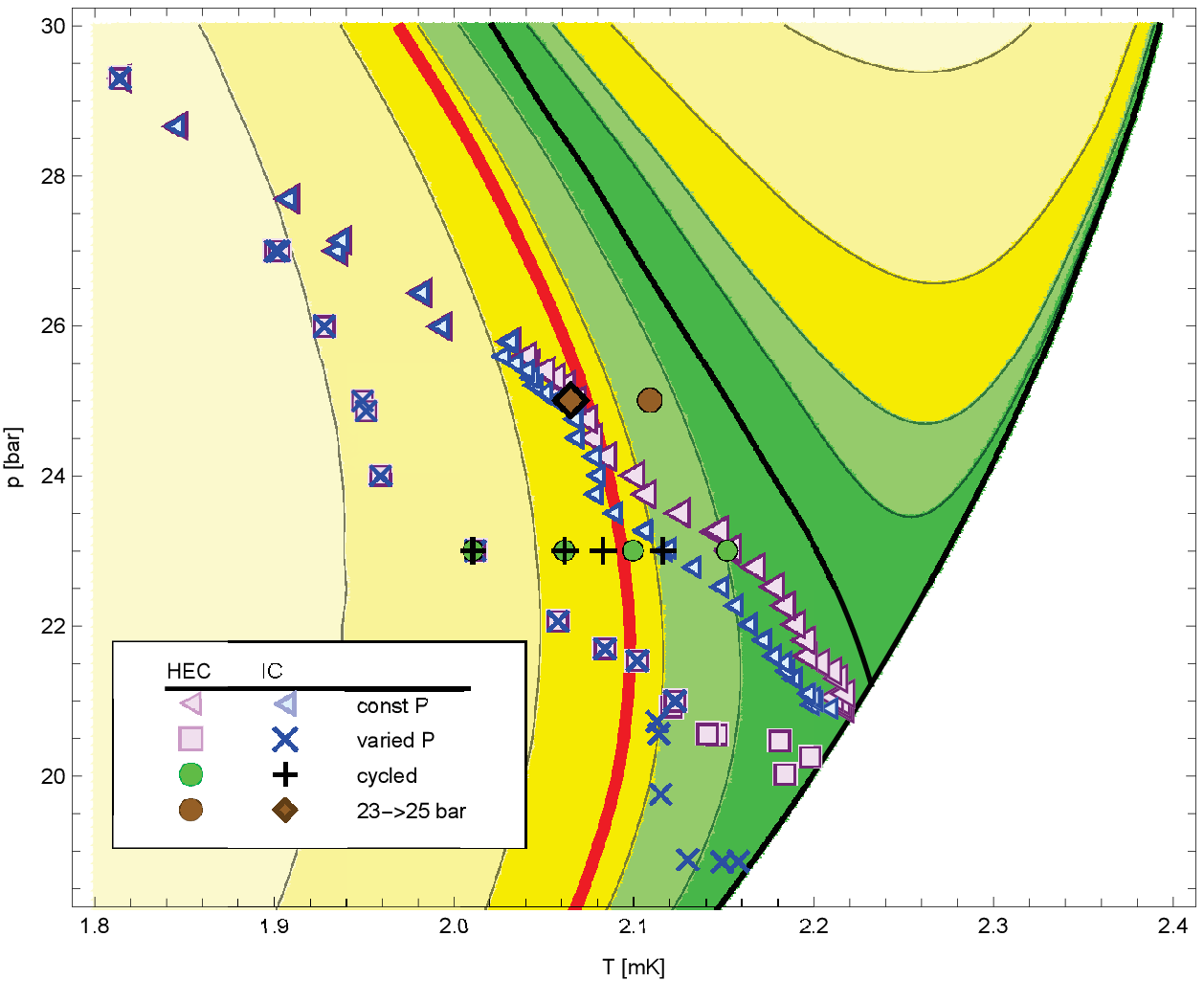}\\
\includegraphics[width=0.60\textwidth]{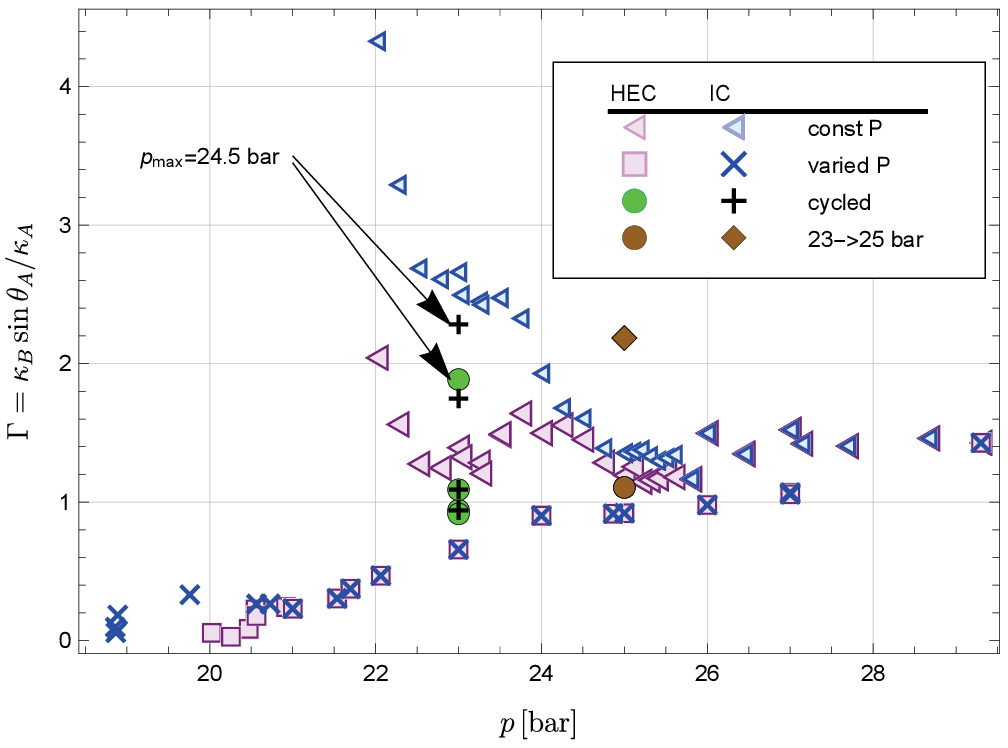}
\caption{
  {\bf  Results from alternate renormalization of strong coupling parameters.}
  Graphs correspond to Figs.~5 and 6 from the main text, but with the alternate model for temperature dependence of the Landau Ginzburg parameters from Supplemental Note \ref{sec::lin}.
 }
\label{fig::lin}
\end{figure}

Figure~\ref{fig::lin}  shows the results of using this alternative approach to modeling the experiment.  While the renormalization scheme causes small changes in the results, our general conclusions are unchanged.   This concordance gives us confidence in our results.

The most notable difference between the renormalization schemes is in the cycled paths, which are cooled at 23 bar, transit to higher pressure, and then continue cooling at 23 bar (shown as green circles and black "+" marks).  In Fig.~\ref{fig::lin} the outliers are somewhat less extreme than in the main text.  The model dependence of $\Gamma$ for these points indicates that one should be cautious about ascribing too much significance to them.

\pagebreak

\begin{thebibliography}{10}
\expandafter\ifx\csname url\endcsname\relax
  \def\url#1{\texttt{#1}}\fi
\expandafter\ifx\csname urlprefix\endcsname\relax\def\urlprefix{URL }\fi
\providecommand{\bibinfo}[2]{#2}
\providecommand{\eprint}[2][]{\url{#2}}

\bibitem{Greywall86SH}
\bibinfo{author}{Greywall, D.}
\newblock \bibinfo{title}{$^{3}$\textrm{He} specific heat and thermometry at
  millikelvin temperatures}.
\newblock \emph{\bibinfo{journal}{Physical Review B}}
  \textbf{\bibinfo{volume}{33}}, \bibinfo{pages}{7520 -- 7538}
  (\bibinfo{year}{1986}).
\newblock \urlprefix\url{http://dx.doi.org/10.1103/PhysRevB.33.7520}.

\bibitem{Wiman2016}
\bibinfo{author}{Wiman, J.~J.} \& \bibinfo{author}{Sauls, J.~A.}
\newblock \bibinfo{title}{Strong-coupling and the stripe phase of
  3-\textrm{He}}.
\newblock \emph{\bibinfo{journal}{Journal of Low Temperature Physics}}
  \textbf{\bibinfo{volume}{184}}, \bibinfo{pages}{1054--1070}
  (\bibinfo{year}{2016}).
\newblock \urlprefix\url{https://doi.org/10.1007/s10909-016-1632-7}.

\bibitem{WheatleyPRL1974}
\bibinfo{author}{Paulson, D.~N.}, \bibinfo{author}{Kojima, H.} \&
  \bibinfo{author}{Wheatley, J.~C.}
\newblock \bibinfo{title}{Profound effect of a magnetic field on the phase
  diagram of superfluid $^{3}\mathrm{He}$}.
\newblock \emph{\bibinfo{journal}{Phys. Rev. Lett.}}
  \textbf{\bibinfo{volume}{32}}, \bibinfo{pages}{1098--1101}
  (\bibinfo{year}{1974}).
\newblock \urlprefix\url{https://link.aps.org/doi/10.1103/PhysRevLett.32.1098}.

\bibitem{LotnykPRL2021}
\bibinfo{author}{Lotnyk, D.} \emph{et~al.}
\newblock \bibinfo{title}{Path-dependent supercooling of the $^{3}\mathrm{He}$
  superfluid $a\ensuremath{-}b$ transition}.
\newblock \emph{\bibinfo{journal}{Phys. Rev. Lett.}}
  \textbf{\bibinfo{volume}{126}}, \bibinfo{pages}{215301}
  (\bibinfo{year}{2021}).
\newblock
  \urlprefix\url{https://link.aps.org/doi/10.1103/PhysRevLett.126.215301}.

\bibitem{Wheatley1974}
\bibinfo{author}{Kleinberg, R.~L.}, \bibinfo{author}{Paulson, D.~N.},
  \bibinfo{author}{Webb, R.~A.} \& \bibinfo{author}{Wheatley, J.~C.}
\newblock \bibinfo{title}{\textrm{Supercooling and superheating of the AB
  transition in superfluid $^3$He near the polycritical point}}.
\newblock \emph{\bibinfo{journal}{J. Low Temp. Phys.}}
  \textbf{\bibinfo{volume}{17}}, \bibinfo{pages}{521--528}
  (\bibinfo{year}{1974}).
\newblock \urlprefix\url{https://doi.org/10.1007/BF00655071}.

\bibitem{HakonenPRL1985}
\bibinfo{author}{Hakonen, P.~J.}, \bibinfo{author}{Krusius, M.},
  \bibinfo{author}{Salomaa, M.~M.} \& \bibinfo{author}{Simola, J.~T.}
\newblock \bibinfo{title}{Comment on ``$\mathrm{N}$ucleation of
  $^{3}\mathrm{He}$-$\mathrm{B}$ from the $\mathrm{A}$ phase: A cosmic-ray
  effect?"}.
\newblock \emph{\bibinfo{journal}{Phys. Rev. Lett.}}
  \textbf{\bibinfo{volume}{54}}, \bibinfo{pages}{245--245}
  (\bibinfo{year}{1985}).
\newblock \urlprefix\url{https://link.aps.org/doi/10.1103/PhysRevLett.54.245}.

\bibitem{Fukuyama1987}
\bibinfo{author}{Fukuyama, H.}, \bibinfo{author}{Ishimoto, H.},
  \bibinfo{author}{Tazaki, T.} \& \bibinfo{author}{Ogawa, S.}
\newblock \bibinfo{title}{$^{3}\mathrm{He}$ melting curve below 15 mk}.
\newblock \emph{\bibinfo{journal}{Phys. Rev. B}} \textbf{\bibinfo{volume}{36}},
  \bibinfo{pages}{8921--8924} (\bibinfo{year}{1987}).
\newblock \urlprefix\url{https://link.aps.org/doi/10.1103/PhysRevB.36.8921}.

\bibitem{Swift1987}
\bibinfo{author}{Swift, G.~W.} \& \bibinfo{author}{Buchanan, D.~S.}
\newblock \bibinfo{title}{{Nucleation and Growth of $^3$He-B in $^3$He-A}}.
\newblock \emph{\bibinfo{journal}{Jpn. J. Appl. Phys.}}
  \textbf{\bibinfo{volume}{26}}, \bibinfo{pages}{1828} (\bibinfo{year}{1987}).
\newblock \urlprefix\url{https://doi.org/10.7567/jjaps.26s3.1828}.

\bibitem{SchifferPRL1992}
\bibinfo{author}{Schiffer, P.}, \bibinfo{author}{O'Keefe, M.~T.},
  \bibinfo{author}{Hildreth, M.~D.}, \bibinfo{author}{Fukuyama, H.} \&
  \bibinfo{author}{Osheroff, D.~D.}
\newblock \bibinfo{title}{Strong supercooling and stimulation of the
  $\mathrm{A-B}$ transition in superfluid $^{3}\mathrm{He}$}.
\newblock \emph{\bibinfo{journal}{Phys. Rev. Lett.}}
  \textbf{\bibinfo{volume}{69}}, \bibinfo{pages}{120--123}
  (\bibinfo{year}{1992}).
\newblock \urlprefix\url{https://link.aps.org/doi/10.1103/PhysRevLett.69.120}.

\bibitem{OsheroffPRL1992}
\bibinfo{author}{Schiffer, P.}, \bibinfo{author}{O'Keefe, M.~T.},
  \bibinfo{author}{Hildreth, M.~D.}, \bibinfo{author}{Fukuyama, H.} \&
  \bibinfo{author}{Osheroff, D.~D.}
\newblock \bibinfo{title}{Strong supercooling and stimulation of the a-b
  transition in superfluid $^{3}\mathrm{He}$}.
\newblock \emph{\bibinfo{journal}{Phys. Rev. Lett.}}
  \textbf{\bibinfo{volume}{69}}, \bibinfo{pages}{120--123}
  (\bibinfo{year}{1992}).
\newblock \urlprefix\url{https://link.aps.org/doi/10.1103/PhysRevLett.69.120}.

\bibitem{Cahn-Hilliard1958}
\bibinfo{author}{Cahn, J.~W.} \& \bibinfo{author}{Hilliard, J.~E.}
\newblock \bibinfo{title}{Free energy of a nonuniform system. i. interfacial
  free energy}.
\newblock \emph{\bibinfo{journal}{J. Chem. Phys.}}
  \textbf{\bibinfo{volume}{28}}, \bibinfo{pages}{258--267}
  (\bibinfo{year}{1958}).
\newblock \urlprefix\url{https://doi.org/10.1063/1.1744102}.

\bibitem{Langer:1969bc}
\bibinfo{author}{Langer, J.~S.}
\newblock \bibinfo{title}{{Statistical theory of the decay of metastable
  states}}.
\newblock \emph{\bibinfo{journal}{Annals Phys.}} \textbf{\bibinfo{volume}{54}},
  \bibinfo{pages}{258--275} (\bibinfo{year}{1969}).

\bibitem{OsheroffPRL1977}
\bibinfo{author}{Osheroff, D.~D.} \& \bibinfo{author}{Cross, M.~C.}
\newblock \bibinfo{title}{Interfacial surface energy between the superfluid
  phases of ${\mathrm{he}}^{3}$}.
\newblock \emph{\bibinfo{journal}{Phys. Rev. Lett.}}
  \textbf{\bibinfo{volume}{38}}, \bibinfo{pages}{905--909}
  (\bibinfo{year}{1977}).
\newblock \urlprefix\url{https://link.aps.org/doi/10.1103/PhysRevLett.38.905}.

\bibitem{kaul1980surface}
\bibinfo{author}{Kaul, R.} \& \bibinfo{author}{Kleinert, H.}
\newblock \bibinfo{title}{Surface energy and textural boundary conditions
  between a and b phases of 3he}.
\newblock \emph{\bibinfo{journal}{Journal of Low Temperature Physics}}
  \textbf{\bibinfo{volume}{38}}, \bibinfo{pages}{539--552}
  (\bibinfo{year}{1980}).
\newblock \urlprefix\url{https://doi.org/10.1007/BF00115487}.

\bibitem{Bailin:1980ny}
\bibinfo{author}{Bailin, D.} \& \bibinfo{author}{Love, A.}
\newblock \bibinfo{title}{{Instantons in Superfluid He-3}}.
\newblock \emph{\bibinfo{journal}{J. Phys. A}} \textbf{\bibinfo{volume}{13}},
  \bibinfo{pages}{L271} (\bibinfo{year}{1980}).

\bibitem{PhysRevLett.53.1096}
\bibinfo{author}{Leggett, A.~J.}
\newblock \bibinfo{title}{Nucleation of $^{3}\mathrm{He}$-$b$ from the $a$
  phase: A cosmic-ray effect?}
\newblock \emph{\bibinfo{journal}{Phys. Rev. Lett.}}
  \textbf{\bibinfo{volume}{53}}, \bibinfo{pages}{1096--1099}
  (\bibinfo{year}{1984}).
\newblock \urlprefix\url{https://link.aps.org/doi/10.1103/PhysRevLett.53.1096}.

\bibitem{Ambegaokar74}
\bibinfo{author}{Ambegaokar, V.}, \bibinfo{author}{deGennes, P.~G.} \&
  \bibinfo{author}{Rainer, D.}
\newblock \bibinfo{title}{Landau-ginsburg equations for an anisotropic
  superfluid}.
\newblock \emph{\bibinfo{journal}{Phys. Rev. A}} \textbf{\bibinfo{volume}{9}},
  \bibinfo{pages}{2676--2685} (\bibinfo{year}{1974}).
\newblock \urlprefix\url{https://link.aps.org/doi/10.1103/PhysRevA.9.2676}.

\bibitem{Wheatley1986}
\bibinfo{author}{Buchanan, D.~S.}, \bibinfo{author}{Swift, G.~W.} \&
  \bibinfo{author}{Wheatley, J.~C.}
\newblock \bibinfo{title}{Velocity of propagation of the $^{3}\mathrm{He}~
  \ensuremath{A-B}$ interface in hypercooled $^{3}\mathrm{He}\ensuremath{-A}$}.
\newblock \emph{\bibinfo{journal}{Phys. Rev. Lett.}}
  \textbf{\bibinfo{volume}{57}}, \bibinfo{pages}{341--344}
  (\bibinfo{year}{1986}).
\newblock \urlprefix\url{https://link.aps.org/doi/10.1103/PhysRevLett.57.341}.

\bibitem{BoydSwift1990}
\bibinfo{author}{Boyd, S. T.~P.} \& \bibinfo{author}{Swift, G.~W.}
\newblock \bibinfo{title}{New mode of growth of $^{3}\mathrm{He}$-$\mathit{B}$
  in hypercooled $^{3}\mathrm{He}$-$\mathit{A}$: Evidence of a spin
  supercurrent}.
\newblock \emph{\bibinfo{journal}{Phys. Rev. Lett.}}
  \textbf{\bibinfo{volume}{64}}, \bibinfo{pages}{894--897}
  (\bibinfo{year}{1990}).
\newblock \urlprefix\url{https://link.aps.org/doi/10.1103/PhysRevLett.64.894}.

\bibitem{BauerleNature1996}
\bibinfo{author}{BÃ€uerle, C.}, \bibinfo{author}{{Yu. M. Bunkov}},
  \bibinfo{author}{Fisher, S.~N.}, \bibinfo{author}{Godfrin, H.} \&
  \bibinfo{author}{Pickett, G.~R.}
\newblock \bibinfo{title}{Laboratory simulation of cosmic string formation in
  the early universe using superfluid $^{3}$\textrm{He}}.
\newblock \emph{\bibinfo{journal}{Nature}} \textbf{\bibinfo{volume}{382}},
  \bibinfo{pages}{332--334} (\bibinfo{year}{1996}).
\newblock \urlprefix\url{https://doi.org/10.1038/382332a0}.

\bibitem{BunkovPRL1998}
\bibinfo{author}{{Yu. M. Bunkov, and O. D. Timofeevskaya}}.
\newblock \bibinfo{title}{``$\mathrm{C}$osmological'' scenario for
  $\mathit{A}\ensuremath{-}\mathit{B}$ phase transition in superfluid
  ${}^{3}\mathrm{He}$}.
\newblock \emph{\bibinfo{journal}{Phys. Rev. Lett.}}
  \textbf{\bibinfo{volume}{80}}, \bibinfo{pages}{4927--4930}
  (\bibinfo{year}{1998}).
\newblock \urlprefix\url{https://link.aps.org/doi/10.1103/PhysRevLett.80.4927}.

\bibitem{BartkowiakPRL2000}
\bibinfo{author}{Bartkowiak, M.} \emph{et~al.}
\newblock \bibinfo{title}{Primary and secondary nucleation of the transition
  between the $\mathit{A}$ and $\mathit{B}$ phases of superfluid
  ${}^{3}\mathrm{He}$}.
\newblock \emph{\bibinfo{journal}{Phys. Rev. Lett.}}
  \textbf{\bibinfo{volume}{85}}, \bibinfo{pages}{4321--4324}
  (\bibinfo{year}{2000}).
\newblock \urlprefix\url{https://link.aps.org/doi/10.1103/PhysRevLett.85.4321}.

\bibitem{LeggettResp1985}
\bibinfo{author}{Leggett, A.~J.}
\newblock \bibinfo{title}{Leggett responds}.
\newblock \emph{\bibinfo{journal}{Phys. Rev. Lett.}}
  \textbf{\bibinfo{volume}{54}}, \bibinfo{pages}{246--246}
  (\bibinfo{year}{1985}).
\newblock \urlprefix\url{https://link.aps.org/doi/10.1103/PhysRevLett.54.246}.

\bibitem{LeggettPRL1986}
\bibinfo{author}{Yip, S.} \& \bibinfo{author}{Leggett, A.~J.}
\newblock \bibinfo{title}{Dynamics of the $^{3}\mathrm{He ~A}\ensuremath{-B}$
  phase boundary}.
\newblock \emph{\bibinfo{journal}{Phys. Rev. Lett.}}
  \textbf{\bibinfo{volume}{57}}, \bibinfo{pages}{345--348}
  (\bibinfo{year}{1986}).
\newblock \urlprefix\url{https://link.aps.org/doi/10.1103/PhysRevLett.57.345}.

\bibitem{LeggettYip1990}
\bibinfo{author}{Leggett, A.~J.} \& \bibinfo{author}{Yip, S.~K.}
\newblock \bibinfo{title}{Nucleation and growth of $^3\mathrm{He-B}$ in the
  supercooled $\mathrm{A}$-phase}.
\newblock In \bibinfo{editor}{Pitaevskii, L.~P.} \& \bibinfo{editor}{Halperin,
  W.~P.} (eds.) \emph{\bibinfo{booktitle}{Helium Three}},
  no.~\bibinfo{number}{26} in \bibinfo{series}{Modern Problems in Condensed
  Matter Sciences}, \bibinfo{type}{chapt.}~\bibinfo{chapter}{8},
  \bibinfo{pages}{523--707} (\bibinfo{publisher}{Elsevier},
  \bibinfo{address}{Amsterdam}, \bibinfo{year}{1990}), \bibinfo{edition}{third}
  edn.

\bibitem{LeggettJLTP}
\bibinfo{author}{Leggett, A.~J.}
\newblock \bibinfo{title}{High-energy low-temperature physics: Production of
  phase transitions and topological defects by energetic particles in
  superfluid $^3\mathrm{He}$}.
\newblock \emph{\bibinfo{journal}{J. Low Temp. Phys.}}
  \textbf{\bibinfo{volume}{126}}, \bibinfo{pages}{775--804}
  (\bibinfo{year}{2002}).
\newblock \urlprefix\url{https://doi.org/10.1023/A:1013878104932}.

\bibitem{HongJLTP}
\bibinfo{author}{Ki~Hong, D.}
\newblock \bibinfo{title}{Q-balls in superfluid $^{3}$\textrm{He}}.
\newblock \emph{\bibinfo{journal}{J. Low Temp. Phys.}}
  \textbf{\bibinfo{volume}{71}}, \bibinfo{pages}{483--494}
  (\bibinfo{year}{1988}).
\newblock \urlprefix\url{https://doi.org/10.1007/BF00116874}.

\bibitem{TyePRB2011}
\bibinfo{author}{Tye, S.-H.~H.} \& \bibinfo{author}{Wohns, D.}
\newblock \bibinfo{title}{Resonant tunneling in superfluid $^{3}$\textrm{He}}.
\newblock \emph{\bibinfo{journal}{Phys. Rev. B}} \textbf{\bibinfo{volume}{84}},
  \bibinfo{pages}{184518} (\bibinfo{year}{2011}).
\newblock \urlprefix\url{https://link.aps.org/doi/10.1103/PhysRevB.84.184518}.

\bibitem{Kirzhnits:1972iw}
\bibinfo{author}{Kirzhnits, D.~A.}
\newblock \bibinfo{title}{{Weinberg model in the hot universe}}.
\newblock \emph{\bibinfo{journal}{JETP Lett.}} \textbf{\bibinfo{volume}{15}},
  \bibinfo{pages}{529--531} (\bibinfo{year}{1972}).
\newblock \bibinfo{note}{[Pisma Zh. Eksp. Teor. Fiz.15,745(1972)]}.

\bibitem{Kirzhnits:1972ut}
\bibinfo{author}{Kirzhnits, D.~A.} \& \bibinfo{author}{Linde, A.~D.}
\newblock \bibinfo{title}{{Macroscopic Consequences of the Weinberg Model}}.
\newblock \emph{\bibinfo{journal}{Phys. Lett.}} \textbf{\bibinfo{volume}{42B}},
  \bibinfo{pages}{471--474} (\bibinfo{year}{1972}).

\bibitem{Kuzmin:1985mm}
\bibinfo{author}{Kuzmin, V.~A.}, \bibinfo{author}{Rubakov, V.~A.} \&
  \bibinfo{author}{Shaposhnikov, M.~E.}
\newblock \bibinfo{title}{{On the Anomalous Electroweak Baryon Number
  Nonconservation in the Early Universe}}.
\newblock \emph{\bibinfo{journal}{Phys. Lett. B}}
  \textbf{\bibinfo{volume}{155}}, \bibinfo{pages}{36} (\bibinfo{year}{1985}).

\bibitem{Witten:1984rs}
\bibinfo{author}{Witten, E.}
\newblock \bibinfo{title}{{Cosmic Separation of Phases}}.
\newblock \emph{\bibinfo{journal}{Phys.Rev.}} \textbf{\bibinfo{volume}{D30}},
  \bibinfo{pages}{272--285} (\bibinfo{year}{1984}).

\bibitem{1986MNRAS.218..629H}
\bibinfo{author}{{Hogan}, C.~J.}
\newblock \bibinfo{title}{{Gravitational radiation from cosmological phase
  transitions}}.
\newblock \emph{\bibinfo{journal}{MNRAS}} \textbf{\bibinfo{volume}{218}},
  \bibinfo{pages}{629--636} (\bibinfo{year}{1986}).

\bibitem{Mazumdar:2018dfl}
\bibinfo{author}{Mazumdar, A.} \& \bibinfo{author}{White, G.}
\newblock \bibinfo{title}{{Review of cosmic phase transitions: their
  significance and experimental signatures}}.
\newblock \emph{\bibinfo{journal}{Rept. Prog. Phys.}}
  \textbf{\bibinfo{volume}{82}}, \bibinfo{pages}{076901}
  (\bibinfo{year}{2019}).
\newblock \eprint{1811.01948}.

\bibitem{Hindmarsh:2020hop}
\bibinfo{author}{Hindmarsh, M.~B.}, \bibinfo{author}{L\"uben, M.},
  \bibinfo{author}{Lumma, J.} \& \bibinfo{author}{Pauly, M.}
\newblock \bibinfo{title}{{Phase transitions in the early universe}}.
\newblock \emph{\bibinfo{journal}{SciPost Phys. Lect. Notes}}
  \textbf{\bibinfo{volume}{24}}, \bibinfo{pages}{1} (\bibinfo{year}{2021}).
\newblock \eprint{2008.09136}.

\bibitem{audley}
\bibinfo{author}{Amaro-Seoane, P.} \emph{et~al.}
\newblock \bibinfo{title}{Laser interferometer space antenna}
  (\bibinfo{year}{2017}).
\newblock \urlprefix\url{https://arxiv.org/abs/1702.00786}.

\bibitem{Caprini:2019egz}
\bibinfo{author}{Caprini, C.} \emph{et~al.}
\newblock \bibinfo{title}{{Detecting gravitational waves from cosmological
  phase transitions with LISA: an update}}.
\newblock \emph{\bibinfo{journal}{JCAP}} \textbf{\bibinfo{volume}{03}},
  \bibinfo{pages}{024} (\bibinfo{year}{2020}).
\newblock \eprint{1910.13125}.

\bibitem{ColemanPRD1977}
\bibinfo{author}{Callan, C.~G.} \& \bibinfo{author}{Coleman, S.}
\newblock \bibinfo{title}{Fate of the false vacuum. ii. first quantum
  corrections}.
\newblock \emph{\bibinfo{journal}{Phys. Rev. D}} \textbf{\bibinfo{volume}{16}},
  \bibinfo{pages}{1762--1768} (\bibinfo{year}{1977}).
\newblock \urlprefix\url{https://link.aps.org/doi/10.1103/PhysRevD.16.1762}.

\bibitem{Linde:1981zj}
\bibinfo{author}{Linde, A.~D.}
\newblock \bibinfo{title}{{Decay of the False Vacuum at Finite Temperature}}.
\newblock \emph{\bibinfo{journal}{Nucl.Phys.}} \textbf{\bibinfo{volume}{B216}},
  \bibinfo{pages}{421} (\bibinfo{year}{1983}).

\bibitem{Zhelev18RSI}
\bibinfo{author}{Zhelev, N.} \emph{et~al.}
\newblock \bibinfo{title}{Fabrication of microfluidic cavities using
  \textrm{S}i-to-glass anodic bonding}.
\newblock \emph{\bibinfo{journal}{Review of Scientific Instruments}}
  \textbf{\bibinfo{volume}{89}}, \bibinfo{pages}{073902}
  (\bibinfo{year}{2018}).
\newblock \urlprefix\url{http://dx.doi.org/10.1063/1.5031837}.

\bibitem{RuutuNature1996}
\bibinfo{author}{Ruutu, V. M.~H.} \emph{et~al.}
\newblock \bibinfo{title}{Vortex formation in neutron-irradiated superfluid 3he
  as an analogue of cosmological defect formation}.
\newblock \emph{\bibinfo{journal}{Nature}} \textbf{\bibinfo{volume}{382}},
  \bibinfo{pages}{334--336} (\bibinfo{year}{1996}).
\newblock \urlprefix\url{https://doi.org/10.1038/382334a0}.

\bibitem{Kibble1976}
\bibinfo{author}{Kibble, T. W.~B.}
\newblock \bibinfo{title}{Topology of cosmic domains and strings}.
\newblock \emph{\bibinfo{journal}{J. Phys. A: Math. Gen.}}
  \textbf{\bibinfo{volume}{9}}, \bibinfo{pages}{1387--1398}
  (\bibinfo{year}{1976}).
\newblock \urlprefix\url{https://doi.org/10.1088}.

\bibitem{Zurek1985}
\bibinfo{author}{Zurek, W.~H.}
\newblock \bibinfo{title}{Cosmological experiments in superfluid helium?}
\newblock \emph{\bibinfo{journal}{Nature}} \textbf{\bibinfo{volume}{317}},
  \bibinfo{pages}{505--508} (\bibinfo{year}{1985}).
\newblock \urlprefix\url{https://doi.org/10.1038/317505a0}.

\bibitem{Maki_JLTP1978}
\bibinfo{author}{Maki, K.}
\newblock \bibinfo{title}{Planar textures in superfluid 3he-a}.
\newblock \emph{\bibinfo{journal}{Journal of Low Temperature Physics}}
  \textbf{\bibinfo{volume}{32}}, \bibinfo{pages}{1--17} (\bibinfo{year}{1978}).
\newblock \urlprefix\url{https://doi.org/10.1007/BF00116903}.

\bibitem{chaikin}
\bibinfo{author}{Chaikin, P.~M.} \& \bibinfo{author}{Lubensky, T.~C.}
\newblock \emph{\bibinfo{title}{Principles of Condensed Matter Physics}}
  (\bibinfo{publisher}{Cambridge University Press}, \bibinfo{year}{1995}).

\bibitem{Greywall84TC}
\bibinfo{author}{Greywall, D.}
\newblock \bibinfo{title}{Thermal conductivity of normal liquid
  $^{3}$\textrm{He}}.
\newblock \emph{\bibinfo{journal}{Physical Review B}}
  \textbf{\bibinfo{volume}{29}}, \bibinfo{pages}{4933 -- 4945}
  (\bibinfo{year}{1984}).
\newblock \urlprefix\url{http://dx.doi.org/10.1103/PhysRevB.29.4933}.

\bibitem{Bartkowiak2004}
\bibinfo{author}{Bartkowiak, M.} \emph{et~al.}
\newblock \bibinfo{title}{Interfacial energy of the superfluid
  $^{3}\mathrm{H}\mathrm{e}$ $a\mathrm{\text{\ensuremath{-}}}b$ phase interface
  in the zero-temperature limit}.
\newblock \emph{\bibinfo{journal}{Phys. Rev. Lett.}}
  \textbf{\bibinfo{volume}{93}}, \bibinfo{pages}{045301}
  (\bibinfo{year}{2004}).
\newblock
  \urlprefix\url{https://link.aps.org/doi/10.1103/PhysRevLett.93.045301}.

\bibitem{Kaul1980}
\bibinfo{author}{Kaul, R.} \& \bibinfo{author}{Kleinert, H.}
\newblock \bibinfo{title}{Surface energy and textural boundary conditions
  between a and b phases of 3 he}.
\newblock \emph{\bibinfo{journal}{Journal of Low Temperature Physics}}
  \textbf{\bibinfo{volume}{38}}, \bibinfo{pages}{539--552}
  (\bibinfo{year}{1980}).

\bibitem{Schopohl1987}
\bibinfo{author}{Schopohl, N.}
\newblock \bibinfo{title}{Spatial dependence of the order parameter of
  superfluid $^{3}\mathrm{He}$ at the a-b phase boundary}.
\newblock \emph{\bibinfo{journal}{Phys. Rev. Lett.}}
  \textbf{\bibinfo{volume}{58}}, \bibinfo{pages}{1664--1667}
  (\bibinfo{year}{1987}).
\newblock \urlprefix\url{https://link.aps.org/doi/10.1103/PhysRevLett.58.1664}.

\bibitem{Thuneberg1991}
\bibinfo{author}{Thuneberg, E.~V.}
\newblock \bibinfo{title}{A-b interface of superfluid $^{3}\mathrm{He}$ in a
  magnetic field}.
\newblock \emph{\bibinfo{journal}{Phys. Rev. B}} \textbf{\bibinfo{volume}{44}},
  \bibinfo{pages}{9685--9691} (\bibinfo{year}{1991}).
\newblock \urlprefix\url{https://link.aps.org/doi/10.1103/PhysRevB.44.9685}.

\bibitem{Balibar2000}
\bibinfo{author}{Balibar, S.}, \bibinfo{author}{Mizusaki, T.} \&
  \bibinfo{author}{Sasaki, Y.}
\newblock \bibinfo{title}{Comments on heterogeneous nucleation in helium}.
\newblock \emph{\bibinfo{journal}{J. Low Temp. Phys.}}
  \textbf{\bibinfo{volume}{120}}, \bibinfo{pages}{293--314}
  (\bibinfo{year}{2000}).

\bibitem{Zhelev17NC}
\bibinfo{author}{Zhelev, N.} \emph{et~al.}
\newblock \bibinfo{title}{The \textrm{A-B} transition in superfluid helium-3
  under confinement in a thin slab geometry}.
\newblock \emph{\bibinfo{journal}{Nature Communications}}
  \textbf{\bibinfo{volume}{8}}, \bibinfo{pages}{15963} (\bibinfo{year}{2017}).
\newblock \urlprefix\url{http://dx.doi.org/10.1038/ncomms15963}.

\bibitem{Hiscock:1995ma}
\bibinfo{author}{Hiscock, W.~A.}
\newblock \bibinfo{title}{{Nucleation of vacuum phase transitions by
  topological defects}}.
\newblock \emph{\bibinfo{journal}{Phys. Lett. B}}
  \textbf{\bibinfo{volume}{366}}, \bibinfo{pages}{77--81}
  (\bibinfo{year}{1996}).
\newblock \eprint{gr-qc/9510003}.

\bibitem{Gregory:2013hja}
\bibinfo{author}{Gregory, R.}, \bibinfo{author}{Moss, I.~G.} \&
  \bibinfo{author}{Withers, B.}
\newblock \bibinfo{title}{{Black holes as bubble nucleation sites}}.
\newblock \emph{\bibinfo{journal}{JHEP}} \textbf{\bibinfo{volume}{03}},
  \bibinfo{pages}{081} (\bibinfo{year}{2014}).
\newblock \eprint{1401.0017}.

\end{thebibliography}

\begin{thebibliography}{10}

\expandafter\ifx\csname url\endcsname\relax
  \def\url#1{\texttt{#1}}\fi
\expandafter\ifx\csname urlprefix\endcsname\relax\def\urlprefix{URL }\fi
\providecommand{\bibinfo}[2]{#2}
\providecommand{\eprint}[2][]{\url{#2}}

\bibitem{Zhelev18RSI}
\bibinfo{author}{Zhelev, N.} \emph{et~al.}
\newblock \bibinfo{title}{Fabrication of microfluidic cavities using
  \textrm{S}i-to-glass anodic bonding}.
\newblock \emph{\bibinfo{journal}{Rev. Sci. Instrum.}}
  \textbf{\bibinfo{volume}{89}}, \bibinfo{pages}{073902}
  (\bibinfo{year}{2018}).
\newblock \urlprefix\url{http://dx.doi.org/10.1063/1.5031837}.

\bibitem{Vollhardt2003}
\bibinfo{author}{Vollhardt, D.} \& \bibinfo{author}{Woelfle, P.}
\newblock \emph{\bibinfo{title}{The Superfluid Phases Of Helium 3}}
  (\bibinfo{publisher}{Dover, Mineola New York}, \bibinfo{year}{2013}).
\newblock \urlprefix\url{https://doi.org/10.1201/b12808}.

\bibitem{wiman1}
\bibinfo{author}{Wiman, J.~J.} \& \bibinfo{author}{Sauls, J.~A.}
\newblock \bibinfo{title}{Superfluid phases of $^{3}\mathrm{He}$ in nanoscale
  channels}.
\newblock \emph{\bibinfo{journal}{Phys. Rev. B}} \textbf{\bibinfo{volume}{92}},
  \bibinfo{pages}{144515} (\bibinfo{year}{2015}).
\newblock \urlprefix\url{https://link.aps.org/doi/10.1103/PhysRevB.92.144515}.

\bibitem{wiman2016}
\bibinfo{author}{Wiman, J.~J.} \& \bibinfo{author}{Sauls, J.~A.}
\newblock \bibinfo{title}{Strong-coupling and the stripe phase of
  3-\textrm{He}}.
\newblock \emph{\bibinfo{journal}{J. Low Temp. Phys.}}
  \textbf{\bibinfo{volume}{184}}, \bibinfo{pages}{1054--1070}
  (\bibinfo{year}{2016}).
\newblock \urlprefix\url{https://doi.org/10.1007/s10909-016-1632-7}.

\bibitem{ChoiPRB07}
\bibinfo{author}{Choi, H.}, \bibinfo{author}{Davis, J.~P.},
  \bibinfo{author}{Pollanen, J.}, \bibinfo{author}{Haard, T.~M.} \&
  \bibinfo{author}{Halperin, W.~P.}
\newblock \bibinfo{title}{Strong coupling corrections to the ginzburg-landau
  theory of superfluid $^{3}\mathrm{He}$}.
\newblock \emph{\bibinfo{journal}{Phys. Rev. B}} \textbf{\bibinfo{volume}{75}},
  \bibinfo{pages}{174503} (\bibinfo{year}{2007}).
\newblock \urlprefix\url{https://link.aps.org/doi/10.1103/PhysRevB.75.174503}.


\bibitem{Greywall86SH}
\bibinfo{author}{Greywall, D.}
\newblock \bibinfo{title}{$^{3}$\textrm{He} specific heat and thermometry at
  millikelvin temperatures}.
\newblock \emph{\bibinfo{journal}{Phys. Rev. B}} \textbf{\bibinfo{volume}{33}},
  \bibinfo{pages}{7520--7538} (\bibinfo{year}{1986}).
\newblock \urlprefix\url{http://dx.doi.org/10.1103/PhysRevB.33.7520}.

\bibitem{PLTS}
\bibinfo{author}{Tian, Y.}, \bibinfo{author}{Smith, E.~N.} \&
  \bibinfo{author}{Parpia, J.}
\newblock \bibinfo{title}{Conversion between $^3\mathrm{He}$ melting curve
  scales below 100 m$\mathrm{K}$}.
\newblock \emph{\bibinfo{journal}{Journal of Low Temperature Physics}}
  \textbf{\bibinfo{volume}{184}}, \bibinfo{pages}{1573--7357}
  (\bibinfo{year}{2022}).
\newblock \urlprefix\url{https://doi.org/10.1007/s10909-022-02721-z}.

\bibitem{regan}
\bibinfo{author}{Regan, R.~C.}, \bibinfo{author}{Wiman, J.~J.} \&
  \bibinfo{author}{Sauls, J.~A.}
\newblock \bibinfo{title}{Vortex phase diagram of rotating superfluid
  $^{3}\mathrm{He}\text{\ensuremath{-}}\mathrm{B}$}.
\newblock \emph{\bibinfo{journal}{Phys. Rev. B}}
  \textbf{\bibinfo{volume}{101}}, \bibinfo{pages}{024517}
  (\bibinfo{year}{2020}).
\newblock \urlprefix\url{https://link.aps.org/doi/10.1103/PhysRevB.101.024517}.

\bibitem{ambegaokar}
\bibinfo{author}{Ambegaokar, V.}, \bibinfo{author}{deGennes, P.~G.} \&
  \bibinfo{author}{Rainer, D.}
\newblock \bibinfo{title}{Landau-ginsburg equations for an anisotropic
  superfluid}.
\newblock \emph{\bibinfo{journal}{Phys. Rev. A}} \textbf{\bibinfo{volume}{9}},
  \bibinfo{pages}{2676--2685} (\bibinfo{year}{1974}).
\newblock \urlprefix\url{https://link.aps.org/doi/10.1103/PhysRevA.9.2676}.

\bibitem{saulssurface}
\bibinfo{author}{Sauls, J.~A.}
\newblock \bibinfo{title}{Surface states, edge currents, and the angular
  momentum of chiral $p$-wave superfluids}.
\newblock \emph{\bibinfo{journal}{Phys. Rev. B}} \textbf{\bibinfo{volume}{84}},
  \bibinfo{pages}{214509} (\bibinfo{year}{2011}).
\newblock \urlprefix\url{https://link.aps.org/doi/10.1103/PhysRevB.84.214509}.

\bibitem{LeggettYip1990}
\bibinfo{author}{Leggett, A.~J.} \& \bibinfo{author}{Yip, S.~K.}
\newblock \bibinfo{title}{Nucleation and growth of $^3\mathrm{He-B}$ in the
  supercooled $\mathrm{A}$-phase}.
\newblock In \bibinfo{editor}{Pitaevskii, L.~P.} \& \bibinfo{editor}{Halperin,
  W.~P.} (eds.) \emph{\bibinfo{booktitle}{Helium Three}},
  no.~\bibinfo{number}{26} in \bibinfo{series}{Modern Problems in Condensed
  Matter Sciences}, \bibinfo{type}{chapt.}~\bibinfo{chapter}{8},
  \bibinfo{pages}{523--707} (\bibinfo{publisher}{Elsevier},
  \bibinfo{address}{Amsterdam}, \bibinfo{year}{1990}), \bibinfo{edition}{third}
  edn.

\bibitem{Viljas2003}
\bibinfo{author}{Viljas, J.} \& \bibinfo{author}{Thuneberg, E.}
\newblock \bibinfo{title}{Stability of a{\textendash}b phase boundary in a
  constriction}.
\newblock \emph{\bibinfo{journal}{Physica B: Condensed Matter}}
  \textbf{\bibinfo{volume}{329-333}}, \bibinfo{pages}{86--87}
  (\bibinfo{year}{2003}).
\newblock \urlprefix\url{https://doi.org/10.1016/s0921-4526(02)01889-6}.

\bibitem{soap}
\bibinfo{author}{Isenberg, C.}
\newblock \emph{\bibinfo{title}{The science of soap films and soap bubbles}}
  (\bibinfo{publisher}{Dover Publications}, \bibinfo{address}{Mineola, NY},
  \bibinfo{year}{1992}).

\bibitem{Bartkowiak2002}
\bibinfo{author}{Bartkowiak, M.} \emph{et~al.}
\newblock \bibinfo{title}{The unique superfluid 3he a-b interface: Surface
  tension and contact angle}.
\newblock \emph{\bibinfo{journal}{Journal of Low Temperature Physics}}
  \textbf{\bibinfo{volume}{126}}, \bibinfo{pages}{533--538}
  (\bibinfo{year}{2002}).
\newblock \urlprefix\url{https://doi.org/10.1023/a:1013758916085}.

\bibitem{Bartkowiak2003}
\bibinfo{author}{Bartkowiak, M.} \emph{et~al.}
\newblock \bibinfo{title}{Superfluid a{\textendash}b surface tension}.
\newblock \emph{\bibinfo{journal}{Physica B: Condensed Matter}}
  \textbf{\bibinfo{volume}{329-333}}, \bibinfo{pages}{122--125}
  (\bibinfo{year}{2003}).
\newblock \urlprefix\url{https://doi.org/10.1016/s0921-4526(02)01919-1}.

\bibitem{Bartkowiak2004}
\bibinfo{author}{Bartkowiak, M.} \emph{et~al.}
\newblock \bibinfo{title}{Interfacial energy of the superfluid
  $^{3}\mathrm{H}\mathrm{e}$ $a\mathrm{\text{\ensuremath{-}}}b$ phase interface
  in the zero-temperature limit}.
\newblock \emph{\bibinfo{journal}{Phys. Rev. Lett.}}
  \textbf{\bibinfo{volume}{93}}, \bibinfo{pages}{045301}
  (\bibinfo{year}{2004}).
\newblock
  \urlprefix\url{https://link.aps.org/doi/10.1103/PhysRevLett.93.045301}.

\bibitem{Kaul1980}
\bibinfo{author}{Kaul, R.} \& \bibinfo{author}{Kleinert, H.}
\newblock \bibinfo{title}{Surface energy and textural boundary conditions
  between a and b phases of 3 he}.
\newblock \emph{\bibinfo{journal}{Journal of Low Temperature Physics}}
  \textbf{\bibinfo{volume}{38}}, \bibinfo{pages}{539--552}
  (\bibinfo{year}{1980}).

\bibitem{Schopohl1987}
\bibinfo{author}{Schopohl, N.}
\newblock \bibinfo{title}{Spatial dependence of the order parameter of
  superfluid $^{3}\mathrm{He}$ at the a-b phase boundary}.
\newblock \emph{\bibinfo{journal}{Phys. Rev. Lett.}}
  \textbf{\bibinfo{volume}{58}}, \bibinfo{pages}{1664--1667}
  (\bibinfo{year}{1987}).
\newblock \urlprefix\url{https://link.aps.org/doi/10.1103/PhysRevLett.58.1664}.

\bibitem{Thuneberg1991}
\bibinfo{author}{Thuneberg, E.~V.}
\newblock \bibinfo{title}{A-b interface of superfluid $^{3}\mathrm{He}$ in a
  magnetic field}.
\newblock \emph{\bibinfo{journal}{Phys. Rev. B}} \textbf{\bibinfo{volume}{44}},
  \bibinfo{pages}{9685--9691} (\bibinfo{year}{1991}).
\newblock \urlprefix\url{https://link.aps.org/doi/10.1103/PhysRevB.44.9685}.

\bibitem{heikkinen2019}
\bibinfo{author}{Heikkinen, P.~J.} \emph{et~al.}
\newblock \bibinfo{title}{Fragility of surface states in topological superfluid
  $^3$\textrm{He}}.
\newblock \emph{\bibinfo{journal}{Nat. Commun.}} \textbf{\bibinfo{volume}{12}},
  \bibinfo{pages}{1574} (\bibinfo{year}{2021}).
\newblock \urlprefix\url{https://doi.org/10.1038/s41467-021-21831-y}.

\end{thebibliography}
{\bf Supplement References:}

\providecommand{\noopsort}[1]{}\providecommand{\singleletter}[1]{#1}%

\end{document}